\documentclass[11pt,a4paper]{article}
\pdfoutput=1
\usepackage{graphicx}
\usepackage{amsfonts}
\usepackage[margin=2cm]{geometry}
\usepackage{amsmath,amssymb,amsthm,array,booktabs,cite,dsfont,mathtools,multirow,placeins,rotating,slashed,stmaryrd,tensor,verbatim}
\usepackage[bookmarksnumbered,linktocpage,pdfstartview=FitH]{hyperref}
\usepackage{dcolumn}
\usepackage[all]{hypcap}
\usepackage[caption=false]{subfig}

\usepackage{bm}% bold math
\usepackage{cancel}
\usepackage{dcolumn}% Align table columns on decimal point

\newcommand{\R}{\mathds{R}}
\newcommand{\C}{\mathds{C}}
\newcommand{\Q}{\mathds{H}}
\newcommand{\Oct}{\mathds{O}}
\newcommand{\Al}{\mathds{A}}
\newcommand{\id}{\mathds{1}}

\newcommand{\sigmabar}{\bar{\sigma}}
\newcommand{\Chi}{\mathcal{X}}

\newcommand{\gamo}[3]{\Gamma^{#1}_{{#2}{#3}}}
\newcommand{\bgamo}[3]{\bar\Gamma^{#1}_{{#2}{#3}}}
\newcommand{\gamquuu}[3]{\Gamma^{\hat{#1}}_{{\hat{#2}}{\hat{#3}}}}

\newcommand{\gamqudd}[3]{\Gamma^{\hat{#1}}_{{\check{#2}}{\check{#3}}}}

\newcommand{\gamqdud}[3]{\Gamma^{\check{#1}}_{{\hat{#2}}{\check{#3}}}}
\newcommand{\gamqddu}[3]{\Gamma^{\check{#1}}_{{\check{#2}}{\hat{#3}}}}
\newcommand{\gamtud}[2]{\Gamma^{3}_{{\hat{#1}}{\check{#2}}}}
\newcommand{\gamtdu}[2]{\Gamma^{3}_{{\check{#1}}{\hat{#2}}}}

\newcommand{\eh}[1]{e_{\hat{#1}}}
\newcommand{\ef}[1]{e_{\check{#1}}}
\newcommand{\ed}[1]{e_{\underline{#1}}}
\newcommand{\eu}[1]{e_{\bar{#1}}}

\newcolumntype{L}[1]{>{\raggedright\let\newline\\\arraybackslash\hspace{0pt}}m{#1}}
\newcolumntype{C}[1]{>{\centering\let\newline\\\arraybackslash\hspace{0pt}}m{#1}}
\newcolumntype{R}[1]{>{\raggedleft\let\newline\\\arraybackslash\hspace{0pt}}m{#1}}

\DeclareMathOperator{\tr}{tr}
\DeclareMathOperator{\Tr}{Tr} % for partial traces only

\DeclareMathOperator{\SO}{SO}

\DeclareMathOperator{\SL}{SL}
\DeclareMathOperator{\SU}{SU}
\DeclareMathOperator{\Sp}{Sp}

\DeclareMathOperator{\Un}{U}
\DeclareMathOperator{\Spin}{Spin}

\newcommand{\be}{\begin{equation}}
\newcommand{\ee}{\end{equation}}

\DeclareMathOperator{\Span}{span}

\newcommand{\ahat}{\hat{a}}
\newcommand{\bhat}{\hat{b}}
\newcommand{\chat}{\hat{c}}

\newcommand{\av}{\check{a}}
\newcommand{\bv}{\check{b}}

\newcommand{\eps}{\varepsilon}

\newcommand{\ibar}{\bar{i}}
\newcommand{\jbar}{\bar{j}}

\newcommand{\susy}{\mathcal{N}}

\begin{document}

\begin{titlepage}%1
\begin{center}
\hfill Imperial/TP/2013/mjd/02\\

\vskip 2cm

{\Huge \bf Super Yang-Mills, division algebras and triality}

\vskip 1.5cm

{\bf A. Anastasiou, L.~Borsten, M.~J.~Duff, L.~J.~Hughes and
S.~Nagy}

\vskip 20pt

{\it Theoretical Physics, Blackett Laboratory, Imperial College London,\\
 London SW7 2AZ, United Kingdom}\\\vskip 5pt
\texttt{alexandros.anastasiou07@imperial.ac.uk}\\
\texttt{leron.borsten@imperial.ac.uk}\\
\texttt{m.duff@imperial.ac.uk}\\
\texttt{leo.hughes07@imperial.ac.uk}\\
\texttt{s.nagy11@imperial.ac.uk}

\end{center}

\vskip 2.2cm

\begin{center} {\bf ABSTRACT}\\[3ex]\end{center}

We give a unified division algebraic description of ($D=3$, $\mathcal{N}=1,2,4,8$), ($D=4$, $\mathcal{N}=1,2,4$), ($D=6$, $\mathcal{N}=1,2$) and ($D=10$, $\mathcal{N}=1$) super Yang-Mills theories. A given $(D=n+2, \mathcal{N})$ theory is completely specified  by selecting a pair $(\Al_n, \Al_{n\susy})$ of division algebras, $\Al_n\subseteq\Al_{n\susy} =\mathds{R},\mathds{C},\mathds{H},\mathds{O}$, where the subscripts denote the dimension of the algebras. We present a master Lagrangian, defined over $\Al_{n\susy}$-valued fields, which encapsulates all cases. Each possibility is obtained from the unique $(\Oct, \Oct)$ ($D=10$, $\mathcal{N}=1$) theory by a combination of Cayley-Dickson halving, which amounts to dimensional reduction, and removing points, lines and quadrangles of the Fano plane, which amounts to consistent truncation. The so-called triality algebras associated with the division algebras allow for a novel formula for the overall (spacetime plus internal) symmetries of the on-shell degrees of freedom of the theories. We use imaginary $\Al_{n\susy}$-valued auxiliary fields to close the non-maximal supersymmetry algebra off-shell. The failure to close for maximally supersymmetric theories is attributed directly to the non-associativity of the octonions.

\vfill

\end{titlepage}

\newpage \setcounter{page}{1} \numberwithin{equation}{section} \tableofcontents 

\section{Introduction}\label{INTRO}

Over the years the relationship between spacetime, supersymmetry
and the division algebras, $\Al=\R, \C, \Q, \Oct$, has been a  recurring theme. In \cite{Gunaydin:1975mp, Gunaydin:2005zz}   the familiar identification of $D=4$ dimensional Minkowski spacetime with the  quadratic Jordan algebra  $\mathfrak{J}_{2}^{\C}$ of $2\times 2$  complex Hermitian matrices was generalised  using the  cubic Jordan algebras first introduced in \cite{Jordan:1933vh}. From the perspective of the present contribution, the crucial observation contained within  \cite{Gunaydin:1975mp, Gunaydin:2005zz}  is that the reduced structure group of  $\mathfrak{J}_{2}^{\Al}$, linear transformations  preserving the quadratic norm,  is isomorphic to $\Spin(1, 1+\dim \Al)$ for $\Al=\R, \C, \Q, \Oct$, suggesting a natural correspondence with $D=3,4,6,10$ Minkowski spacetime.  
This is reflected by  the Lie algebra isomorphism \cite{Sudbery:1984}
\be
\mathfrak{so}(1,1+n)\cong\mathfrak{sl}(2,\Al),~~~~n=\dim{\Al},
\label{ISO}
\ee
which constitutes an important component of the picture to be developed here.  These relations make it  clear  that the normed division algebras   provide a natural framework for relativistic physics in the critical spacetime dimensions. The same is in fact  true for supersymmetry, as already demonstrated in \cite{Gunaydin:1975mp}. In  particular, the $D=3,4,6,10$ Fierz identities, central to supersymmmetry,   have been shown to follow from the adjoint identities of the Jordan algebra formalism \cite{Sierra:1987}.  Indeed, the super Poincar\'e groups can be  related to the division  algebras, as shown in  \cite{Gursey:1987mv}. The close connections existing between  division algebras, Fierz identities and super Poincar\'e groups led to the important  conclusion  that the classical Green-Schwarz superstring and $\susy=1$ super Yang-Mills (SYM) theories of a single vector and spinor can exist only in the critical dimensions associated with the division algebras \cite{Kugo:1982bn, Green:1987sp}. Moreover, it was shown in \cite{dray2000octonionic,Baez:2009xt} that the supersymmetry in these theories follows directly from the property that multiplication of division algebra elements is \emph{alternative} (to be defined below).  These early observations have since led to numerous developments  intertwining   division algebras, spacetime and supersymmetry. See,  for example, \cite{Gunaydin:1976vq, Gunaydin:1979df, Evans:1987tm, Duff:1987qa,Blencowe:1988sk, Gunaydin:1992zh, Berkovits:1993hx, Evans:1994cn, Schray:1994ur, Manogue:1993ja, Manogue:1998rv, Gunaydin:2000xr, Baez:2001dm,Toppan:2003yx,  Kuznetsova:2006ws, Borsten:2008wd, Baez:2009xt, Baez:2010ye, Gunaydin:2010fi, Rios:2011fa} and the references therein.

In addition to  the evident harmony  of $D=3,4,6,10$ and $\R,\C,\Q,\Oct$, there is also a division algebraic interpretation of the fraction of maximal 
supersymmetries $\nu=\frac{1}{8},\frac{1}{4},\frac{1}{2},1\sim  \R,\C,\Q,\Oct$, corresponding to $0,1,3,7$ lines of the Fano plane. See, for example, \cite{Duff:2006ue,Borsten:2008wd}.  This  complementary role for the division algebras was exploited in \cite{Borsten:2013bp} where we gave an $\R,\C,\Q,\Oct$ description of $\susy=1,2,4,8$ Yang-Mills in $D=3$.

The appearance of the exceptional groups as U-duality symmetries in supergravity  \cite{Cremmer:1979up,deWit:1981eq} and M-theory \cite{Duff:1990hn,Hull:1994ys} also suggests a connection with the octonions, since the octonions offer intuitive descriptions of these groups \cite{Freudenthal:1954,Tits:1955,  Rosenfeld:1956, Freudenthal:1959, Tits:1966, Barton:2003,Baez:2001dm}.  Recently, it was shown \cite{Borsten:2013bp} that  tensoring left and right $D=3, \susy=1,2,4,8$ Yang-Mills multiplets results in sixteen $D=3$ supergravities  with U-dualities filling out  the  magic square of Freudenthal, Rozenfeld and Tits\cite{Freudenthal:1954,Tits:1955,  Rosenfeld:1956, Freudenthal:1959}. The principal aim of the present paper is to develop a division algebraic formulation of $\susy$-extended super-Yang-Mills theories in $D=4,6,10$, as well as $D=3$, a result of interest in its own right. As an extra bonus, however, we will show in a subsequent paper \cite{Anastasiou:2013hba} how tensoring also yields the corresponding $D=4,6,10$ supergravities. In this way we obtain a ``magic pyramid'' of supergravities with the $4 \times 4$ magic square at its base in $D=3$, a $3 \times 3$ square in $D=4$, a $2 \times 2$ square in $D=6$ and Type II supergravity at the apex in $D=10$.

These magic squares of conventional supergravities should not be confused with the important earlier $\susy=2$ ``magic supergravities'' in $D=5,4,3$ constructed in \cite{Gunaydin:1983rk, Gunaydin:1983bi, Gunaydin:1984ak} using cubic Jordan algebras. Their U-duality  groups also appear in the magic square and  correspond to  the symmetries of the generalised Jordan algebraic spacetimes  \cite{Gunaydin:2005zz}. However, their real forms are different from ours as is the number of supersymmetries, which for $D=3$ is given by the sixteen possibilities $\susy=\dim \Al_L + \dim \Al_R$ for $\Al_L, \Al_R=\R, \C, \Q, \Oct$.

Our starting point is an  explicit octonionic formulation $D=10, \susy=1$ super-Yang-Mills theory, the existence of which was suggested by \cite{Kugo:1982bn}. From there we build a unified division algebraic description of ($D=3$, $\mathcal{N}=1,2,4,8$), ($D=4$, $\mathcal{N}=1,2,4$), ($D=6$, $\mathcal{N}=1,2$) and ($D=10$, $\mathcal{N}=1$) SYM theories. Each theory is written in terms of a pair of division algebras: one algebra $\Al_n$ to specify the spacetime dimension $D=n+2$ and another $\Al_{n\susy}$ to specify the number of supersymmetries $\susy$. In our framework, dimensional reduction amounts to `Cayley-Dickson halving', that is, writing an octonion as a pair of quaternions, a quaternion as a pair of complex numbers or a complex number as a pair of real numbers. Starting from ($D=10,$ $\susy=1$), we obtain the maximal theories in $D=3,4,6$. This corresponds to a manifestly octonionic realisation of the maximally supersymmetric Yang-Mills theories in $D=3,4,6,10$. Consistent truncation to theories with lower $\susy$ corresponds to the removal of points, lines and quadrangles of the Fano plane, again emphasising the special role played by the division algebras. Bringing these components together we present a single master Lagrangian; the spacetime dimension,  field content, action and (supersymmetric) transformation rules of each  theory are uniquely determined by specifying the   two division algebras $\Al_{n}, \Al_{n\susy}$ alone. Note, this construction  relies on a generalisation of the well-known identification between 1-forms in $D=2+\dim \Al$ and elements of $\mathfrak{J}_{2}^{\Al}$ to arbitrary $p$-forms.

 We also reveal the important role of triality algebras,  originally appearing in the physics literature in \cite{Gunaydin:1973, Gunaydin:1978},  as the symmetries of the on-shell degrees of freedom of $\susy=1$ SYM theories. By defining a new algebra $\widetilde{\mathfrak{tri}}$, which accounts for both $\Al_n$ and $\Al_{n\susy}$, the on-shell symmetries for theories with any $(D,\susy)$ are summarised in a single formula. 

Finally, we discuss the use of division algebraic auxiliary fields to close the supersymmetry algebra off-shell. This works well for $\nu=\frac{1}{8},\frac{1}{4},\frac{1}{2}$ theories but fails in the maximal $\nu=1$ theories, which are written over $\Oct$. We demonstrate explicitly that this failure to close is a direct result of the non-associativity of the octonions, as hinted at in \cite{Berkovits:1993hx}.

\section{The Normed Division Algebras}\label{NDA}

We begin by discussing the definition of the normed division algebras and some of their properties. An algebra $\Al$ is a vector space equipped with a bilinear multiplication rule and a unit element. We say $\Al$ is a division algebra if, given $x,y\in\Al$ with $xy=0$, then either $x=0$ or $y=0$. A normed division algebra is an algebra $\Al$
equipped with a positive-definite norm satisfying the condition 
\be
|\hspace{-0.2mm}|xy|\hspace{-0.2mm}|=|\hspace{-0.2mm}|x|\hspace{-0.2mm}|\hspace{0.4mm}|\hspace{-0.2mm}|y|\hspace{-0.2mm}|,
\ee
which also implies $\Al$ is a division algebra. From now on it shall be understood that the term `division algebra' is short for `normed division algebra', since we shall have no cause to use division algebras that are not normed.

There is a remarkable theorem due to Hurwitz \cite{Hurwitz:1898}, which states that there are only four normed division algebras: the real numbers $\R$, the complex numbers $\C$,
the quaternions $\Q$ and the octonions $\Oct$. The algebras have dimensions $n=1,2,4$ and $8$, respectively. They can be constructed, one-by-one, using the Cayley-Dickson doubling method, starting with $\R$; the complex numbers are pairs of real numbers equipped with a particular multiplication rule, a quaternion is a pair of complex numbers and an octonion is a pair of quaternions. At the level of vector spaces:
\be
\begin{split}
\C &\cong\R^2,\\
\Q &\cong\C^2 \hspace{0.4mm} \cong\R^4,\\
\Oct &\cong\Q^2\cong\C^4\cong\R^8.
\end{split}
\ee
The real numbers are ordered, commutative and associative, but with each doubling one such property is lost: $\C$ is commutative and associative, $\Q$ is associative, $\Oct$ is \emph{non-associative}. The Cayley-Dickson procedure yields an infinite sequence of algebras, but in doubling the octonions to obtain the 16-dimensional `sedenions' the division algebra property is lost. Sometimes it will be useful to denote the division algebra of dimension $n$ by $\Al_n$. When there is no subscript it is assumed that the division algebra has dimension $n$.

Although the octonions are non-associative, they enjoy the weaker property of \emph{alternativity}. An algebra $\Al$ is alternative if and only if for all $x,y\in\Al$ we have:
\be
(xx)y=x(xy),~~~~~~(xy)x=x(yx),~~~~~~(yx)x=y(xx)
\ee
(note that one of these conditions may be derived from the other two; we write all three just to emphasise the symmetry \cite{Baez:2001dm}). This property is trivially satisfied by the three associative division algebras $\R,\C$ and $\Q$, and so we conclude that the division algebras are alternative. We will see later that this property is crucial for supersymmetry in $D=3,4,6,10$. The three conditions can be neatly summed up if we define a trilinear map called the associator given by:
\be
[x,y,z]=(xy)z-x(yz),~~~~x,y,z\in\Al,
\ee
which measures the failure of associativity. An algebra $\Al$ is then alternative if and only if the associator is an antisymmetric function of its three arguments.

\begin{figure}[h!]
  \centering
    \includegraphics[width=0.3\textwidth]{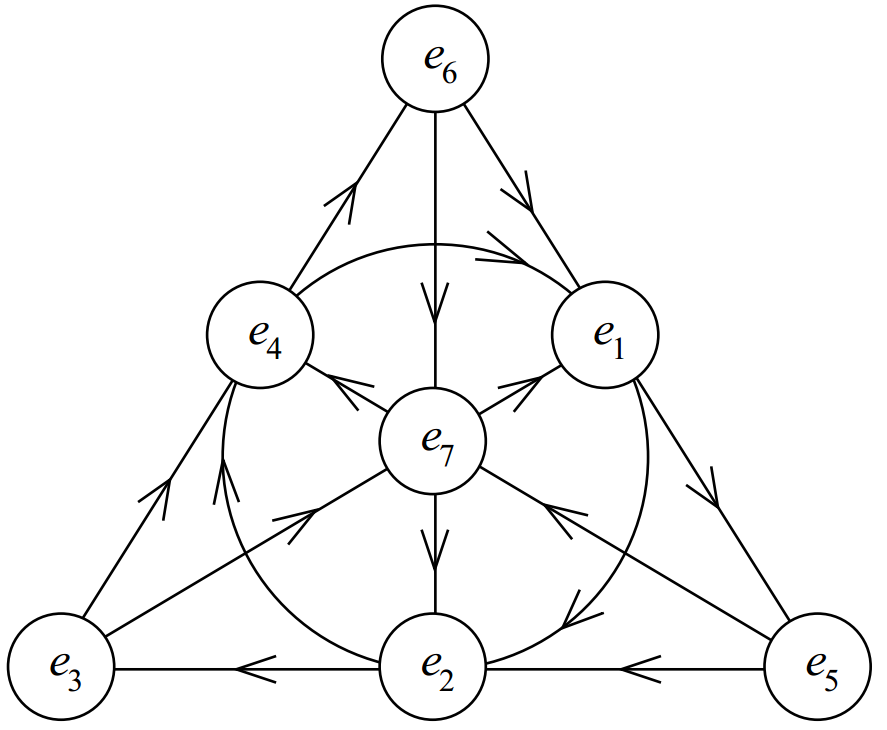}
  \caption{\footnotesize{The Fano plane \cite{Baez:2001dm}. Each oriented line corresponds to a quaternionic subalgebra. For example, $e_2e_3=e_5$ and cyclic permutations; odd permutations go against the direction of the arrows on the Fano plane and we pick up a minus sign, e.g. $e_3e_2=-e_5$.}}\label{FANO}
\end{figure}

A division algebra element $x\in\Al$ is written as the linear combination of $n$ basis elements with real coefficients: $x=x_{a}e_{a}$, with $x_a\in\R$ and $a=0,\cdots,(n-1)$. The first basis element $e_0=1$ is real, while the other $(n-1)$ $e_i$ are imaginary:
\be
e_0^2=1,~~~~e_i^2=-1,
\ee
where $i=1,\cdots,(n-1)$. In analogy with the complex case, we define a conjugation operation indicated by *, which changes the sign of the imaginary basis elements:
\be
{e_0}^*=e_0,~~~~{e_i}^*=-e_i.
\ee
It is natural then to define the real and imaginary parts of $x\in\Al$ by
\be
\text{Re}(x)\equiv\frac{1}{2}(x+x^*)=x_0,~~~~~~\text{Im}(x)\equiv\frac{1}{2}(x-x^*)=x_ie_i.
\ee
Note that this differs slightly with the convention typically used for the complex numbers (since $\text{Im}(a+e_1b)=e_1b$ rather than $b$). The multiplication rule for the basis elements of a general division algebra is given by:
\be\label{OCTMULT}
\begin{split}
&{e_a}{e_b}=\left(+\delta_{a0}\delta_{bc}+\delta_{0b}\delta_{ac}-\delta_{ab}\delta_{0c}+C_{abc}\right)e_c\equiv\gamo{a}{b}{c}e_c,\\
&{e_a^*}{e_b}=\left(+\delta_{a0}\delta_{bc}-\delta_{0b}\delta_{ac}+\delta_{ab}\delta_{0c}-C_{abc}\right)e_c\equiv\bgamo{a}{b}{c}e_c,\\
&{e_a}{e_b^*}=\left(-\delta_{a0}\delta_{bc}+\delta_{0b}\delta_{ac}+\delta_{ab}\delta_{0c}-C_{abc}\right)e_c\equiv\bgamo{c}{a}{b}e_c,
\end{split}
\ee
where we define the structure constants\footnote{The choice of index structure is for later convenience - see equations (\ref{CLIFFORD}), (\ref{PSIMULT}) and (\ref{CLIFFORD2}).}
\be\label{STCON}
\begin{split}
\Gamma^a_{bc}&=\delta_{a0}\delta_{bc}+\delta_{b0}\delta_{ac}-\delta_{ab}\delta_{c0}+C_{abc},\\
\bar\Gamma^a_{bc}&=\delta_{a0}\delta_{bc}-\delta_{b0}\delta_{ac}+\delta_{ab}\delta_{c0}-C_{abc}~~\Rightarrow~~\Gamma^a_{bc}=\bar\Gamma^a_{cb}.
\end{split}
\ee
The tensor $C_{abc}$ is totally antisymmetric with $C_{0ab}=0$, which means all of its components are identically zero for $\Al=\R,\C$. For the quaternions $C_{ijk}$ is simply the permutation symbol $\varepsilon_{ijk}$, while for the octonions the non-zero $C_{ijk}$ are specified by the set $\bold{L}$ of oriented lines of the Fano plane \cite{Baez:2001dm}, which can be used as a mnemonic for octonionic multiplication - see Fig. \ref{FANO}:
\be
\begin{split}
C_{ijk}(\Al) &= \begin{cases} 0 &\mbox{for } \Al=\R,\C \\
1 \hspace{0.1cm}\text{ if }\hspace{0.1cm}ijk=123 & \mbox{for } \Al=\Q  \\
1 \hspace{0.1cm}\text{ if }\hspace{0.1cm}ijk\in\bold{L} & \mbox{for } \Al=\Oct,
\end{cases}\\
\hspace{0.1cm}\text{ where }\hspace{0.1cm}\bold{L}&=\{124,235,346,457,561,672,713\}.
\end{split}
\ee
It is useful to remember that adding 1 (modulo 7) to each of the digits labelling a line in $\bold{L}$ produces the next line. For example, $124\rightarrow235$. 

Restricting to any single line of the Fano plane restricts the octonions to a quaternionic subalgebra so that $C_{ijk}$ reduces to the permutation symbol $\varepsilon_{ijk}$. For example, the subalgebra spanned by $\{e_0,e_1,e_2,e_4\}$ is isomorphic to the quaternions. This will be important for the dimensional reductions carried out in Section \ref{SYMMETRIES}.

The norm $|\hspace{-0.2mm}|x|\hspace{-0.2mm}|$ of a division algebra element $x$ can be defined using the multiplication and conjugation rules as \cite{Baez:2001dm}:
\be
|\hspace{-0.2mm}|x|\hspace{-0.2mm}|^2=xx^*=x^*x=x_a x_a.
\ee
By the polarisation identity we obtain a natural inner product \cite{Baez:2001dm}:
\be
\langle{x}|{y}\rangle=\frac{1}{2}\left(xy^*+yx^*\right)=\frac{1}{2}\left(x^*y+y^*x\right)=x_ay_a\hspace{0.4cm}\text{i.e.}\hspace{0.4cm}\langle{e_a}|{e_b}\rangle=\delta_{ab}.
\ee
This is just the canonical inner product on $\R^n$, which is preserved by $\SO(n)$. This group and its Lie algebra thus have a natural action on the division algebra elements that we will explore in detail later.

There is another Lie algebra associated with the division algebras that will turn out to have physical relevance; we define the triality algebra of $\Al$ as follows:
\be\label{TRIDEF}
\mathfrak{tri}(\mathds{A})\equiv\{(A,B,C)|A(xy)=B(x)y+xC(y)\},\hspace{0.2cm}A,B,C\in\mathfrak{so}(n),\hspace{0.2cm}x,y\in\mathds{A}.
\ee
This algebra appears explicitly in the magic square formula of Barton and Sudbury \cite{Barton:2003}. Although it is not obvious at first sight, the triality algebras turn out to be:
\begin{align}\label{TRI}
&\mathfrak{tri}(\R)=\O,\nonumber\\
&\mathfrak{tri}(\C)=\mathfrak{u}(1)\oplus\mathfrak{u}(1),\nonumber\\
&\mathfrak{tri}(\Q)=\mathfrak{sp}(1)\oplus\mathfrak{sp}(1)\oplus\mathfrak{sp}(1),\nonumber\\
&\mathfrak{tri}(\Oct)=\mathfrak{so}(8).
\end{align}
It will be shown in Section \ref{SYMMETRIES} that the symmetries of the on-shell degrees of freedom of $\mathcal{N}=1$ super Yang-Mills theories in $D=3,4,6,10$ are exactly the triality algebras given above.

Finally, we provide some useful identities for working with octonions and their components. Just as multiplication of the octonionic basis elements is encoded in the lines of the Fano plane, the associator of three octonionic basis elements is encoded in its seven quadrangles\footnote{A quadrangle is the shape we are left with if we remove a line from the Fano plane. Thus the Fano plane has seven points, seven lines and seven quadrangles.} $\bold{Q}$:
\be
[e_a,e_b,e_c]=2Q_{abcd}e_d,
\ee
where the tensor $Q_{abcd}$ is totally antisymmetric with $Q_{0abc}=0$, and the non-zero $Q_{ijkl}$ are given by:
\be
Q_{ijkl}=1\hspace{0.2cm}\text{if}\hspace{0.2cm}ijkl\in\bold{Q}=\{3567,4671,5712,6123,7234,1345,2456\}.
\ee
Since a quadrangle is the complement of a line in the Fano plane, by definition, the tensors $Q_{ijkl}$ and $C_{ijk}$ are dual to one another:
\be
Q_{ijkl}=-\frac{1}{3!}\varepsilon_{ijklmnp}C_{mnp},
\ee
and are also related \cite{dundarer1991octonionic} by
\be
\begin{split}
C_{ijm}C_{klm}\hspace{0.2cm}&=\hspace{0.1cm}\delta_{ik}\delta_{jl}-\delta_{il}\delta_{jk}+Q_{ijkl},\\
C_{ijn}Q_{klmn}\hspace{0.08cm}&=\hspace{0.1cm}3(C_{i[kl}\delta_{m]j}-C_{j[kl}\delta_{m]i}),\\
Q^{ijkl}Q_{mnpl}&=\hspace{0.1cm}6\delta^{[i}_m\delta^j_n\delta^{k]}_p-C^{ijk}C_{mnp}+9Q^{[ij}{}_{[mn}\delta^{k]}_{p]}.
\end{split}
\ee
Note that indices are raised and lowered with the Kronecker delta, so the upper index placement in the final formula is only for notational convenience.

\section{Spacetime Fields in $D=n+2$}\label{SPACE}

In this section we will see how the division algebras can be used to describe field theory in Minkowski space using the Lie algebra isomorphism 
\be
\mathfrak{so}(1,1+n)\cong\mathfrak{sl}(2,\Al).
\ee
We will proceed in direct analogy with the familiar case $\mathfrak{so}(1,3)\cong\mathfrak{sl}(2,\C)$ in $D=4$, showing that the isomorphism allows for the description of spacetime fields and spacetime transformations over the division algebras.
\\
\begin{table}[h!]
\begin{center}
\begin{tabular}{c|c|c|c}
\hline
\hline
\hspace{0.2cm}Field Symbol\hspace{0.2cm} &\hspace{0.2cm}Representation\hspace{0.2cm}
&\hspace{0.2cm} Rep. Symbol\hspace{0.2cm} &\hspace{0.9cm} Group\hspace{0.9cm}
\\
\hline
$\Psi_{\Al}$ & Spinor & $S_+$ & $\SO(1,n+1)$ \\
$\mathcal{X}_{\Al}$ & Conjugate Spinor & $S_-$ & $\SO(1,n+1)$ \\
$A_{\Al}$ & Vector & $V$ & $\SO(1,n+1)$ \\
$\psi_{\Al}$ & Spinor & $s$ & $\SO(n)$ \\
$\chi_{\Al}$ & Conjugate Spinor & $c$ & $\SO(n)$ \\
$a_{\Al}$ & Vector & $v$ & $\SO(n)$ \\
\hline
\hline
\end{tabular}
\end{center}
\caption{\footnotesize{A summary of the fields and notation used in $D=n+2$}}\label{GLOS}
\end{table}
\\
We will first construct the spinor and conjugate spinor representations as they correspond to the fundamental and anti-fundamental representations of $\Spin(1,1+n)$. Then a natural set of generalised Pauli matrices allows a description of vectors and forms analogous to the $D=4$ case. Finally, it will be shown that the on-shell degrees of freedom of massless vector, spinor and conjugate spinor fields can be parametrised each by a single number in $\mathds{A}$.

Table \ref{GLOS} summarises the notation used for the various fields appearing throughout this paper. In relations involving a general division algebra the subscript $\Al$ will be suppressed. In fact, we only use the subscripts in the next section, where fields written over different algebras appear in the same equations. 

\subsection{The Isomorphism $\mathfrak{so}(1,1+n)\cong\mathfrak{sl}(2,\Al)$}\label{ISO}

The isomorphism in question holds for the same reasoning used in $D=4$ for $\mathfrak{so}(1,3)\cong\mathfrak{sl}(2,\C)$. In dimension $D=n+2$ a vector $X$ in spacetime is represented by the components: 
\be
X^\mu=(X^0,X^1,\dots,X^n,X^{n+1})\equiv(t,X^{a+1},z),~~~a=0,1,\ldots,(n-1).
\ee
The vector can also be parametrised \cite{Chung:1987in,Baez:2009xt} by a $2\times2$ Hermitian matrix with entries in $\Al$:
\be
X=\begin{pmatrix}
t+z & x^* \\
x & t-z
\end{pmatrix}\hspace{0.2cm}\text{where}\hspace{0.2cm}t,z\in\R\hspace{0.2cm}\text{and}\hspace{0.2cm}x=X^{a+1}e_a\in\Al.
\ee
Then the determinant of the matrix is the Minkowski metric for $D$-dimensional spacetime:
\be
\det X=t^2-z^2-|x|^2.
\ee
The group of determinant-preserving transformations $\SL(2,\Al)$ then correspond to Lorentz transformations, although care is needed to define elements of this group and its Lie algebra, due to the potential non-commutativity and non-associativity of $\Al$. 

In $D=4$ the Pauli matrices $\{\sigmabar_\mu\}$ are used as a basis for Hermitian matrices, so that we can write $X=X^\mu\bar\sigma_\mu$. This suggests a generalised set of Pauli matrices for $\mu=0,1,\dots,(n+1)$.
The straightforward generalisation of the usual Pauli matrices to all four normed division algebras is the basis \cite{Schray:1994ur,Evans:1994cn}
\be
\begin{split}
\bar\sigma_\mu&=\sigma^\mu=(+\id,\sigma_{a+1},\sigma_{n+1}),\\
\sigma_\mu&=\bar\sigma^\mu=(-\id,\sigma_{a+1},\sigma_{n+1}),
\end{split}
\ee
where
\be
\sigma_{a+1}=\begin{pmatrix}
0 & e_a^* \\
e_a & 0 \end{pmatrix},~~~~~~\sigma_{n+1}=\begin{pmatrix}
1 & 0 \\
0 & -1 \end{pmatrix}.
\ee
The notation is chosen so that in $D=4$ (where $n=2$ and $\Al=\C$) the matrices reduce to the usual Pauli set:
\be
\sigma_1=\begin{pmatrix}
0 & e_0^* \\
e_0 & 0 \end{pmatrix}=\begin{pmatrix}
0 & 1 \\
1 & 0 \end{pmatrix},~~~~~~\sigma_2=\begin{pmatrix}
0 & e_1^* \\
e_1 & 0 \end{pmatrix}\equiv\begin{pmatrix}
0 & -i \\
i & 0 \end{pmatrix},~~~~~~\sigma_3=\begin{pmatrix}
1 & 0 \\
0 & -1 \end{pmatrix}.
\ee
It is easy to see that the generalised Pauli matrices satisfy the required algebra:
\be
\begin{split}
\sigma^\mu\bar\sigma^\nu+\sigma^\nu\bar\sigma^\mu&=2\eta^{\mu\nu}\id,\\
\sigmabar^\mu\sigma^\nu+\sigmabar^\nu\sigma^\mu&=2\eta^{\mu\nu}\id.
\end{split}
\ee
As a result, they can be used to construct Lorentz generators of the spinor and conjugate spinor representations, which we will see explicitly in the following subsection. Note that in $D=3$, the matrices 
\be
\gamma^\mu\equiv\sigma^\mu\eps~~\text{with}~~\varepsilon\equiv\begin{pmatrix}
0 ~ -1 \\
1 ~~~~ 0 \end{pmatrix}
\ee
satisfy the Clifford algebra
\be
\begin{split}
\gamma^\mu\gamma^\nu+\gamma^\nu\gamma^\mu=2\eta^{\mu\nu}\id,
\end{split}
\ee
so using the two sets of matrices $\sigma^\mu$ and $\sigmabar^\mu$ is not strictly necessary. However, we will work with the sigmas for consistency with $D=4,6,10$.

\subsection{Spinors and Pauli Matrices}\label{SPINORS}

As a consequence of the isomorphism above, the spinor $\Psi$ and conjugate spinor $\Chi$ of $\Spin(1,n+1)$ can be written \cite{Chung:1987in,Schray:1994ur,Baez:2009xt} as $2\times1$ columns with entries in $\Al$:
\be
\begin{split}
\Psi&=\begin{pmatrix}
\Psi_1 \\ \Psi_2 \end{pmatrix},~~~~\hspace{0.6mm}\Psi_{1,2}\in\Al,\\
\Chi&=\begin{pmatrix}
\Chi^{\dot1} \\ \Chi^{\dot2}\end{pmatrix},~~~~\Chi^{\dot{1},\dot{2}}\in\Al,
\end{split}
\ee
i.e. $\Psi,\Chi\in{\Al}^2$. We adopt the dotted and un-dotted notation here to distinguish the two representations. However, not all of the usual identities used in $D=4$ hold in the general division algebraic case due to non-commutativity and non-associativity; extra care must be taken to derive universal identities\footnote{In particular, unlike in the $D=4$ case, the spinor $\Phi$, defined by $\Phi=\varepsilon\Psi^*$ with $\varepsilon$ given above, does not transform like a conjugate spinor \cite{Schray:1994fc}. But this will not be a problem in the present paper.}.

We now have the unifying picture that the minimal
spinors in various spacetime dimensions can be obtained from one another simply by switching division algebra
\cite{Baez:2009xt}:
\begin{itemize}
\item{When $\mathds{A}=\mathds{R}$ , $S_{+}\cong{S_-}$ , $\Psi$ is the Majorana
spinor in $D=3$}
\item{When $\mathds{A}=\mathds{C}$ , $S_{+}$ and ${S_-}$ , $\Psi$ and $\Chi$
are the Weyl spinors\footnote{The $S_{+}$ and ${S_-}$ in $D=4$ are related by complex conjugation.} in $D=4$}
\item{When $\mathds{A}=\mathds{H}$ , $S_{+}$ and ${S_-}$ , $\Psi$ and $\Chi$
are the Symplectic-Weyl spinors in $D=6$}
\item{When $\mathds{A}=\mathds{O}$ , $S_{+}$ and ${S_-}$ , $\Psi$ and $\Chi$
are the Majorana-Weyl spinors in $D=10$}.
\end{itemize}
The missing piece of the puzzle is how to transform these representations under $\SL(2,\Al)$. In order to write down an infinitesimal $\mathfrak{sl}(2,\Al)$ transformation of the
spinor and conjugate spinor fields we seek a generalisation of the equations:
\be
\begin{split}\label{LTS1}
&\delta\Psi=\frac{1}{4}\lambda^{\mu\nu}\sigma_{\mu\nu}\Psi=\frac{1}{4}\lambda^{\mu\nu}\sigma_\mu\bar\sigma_\nu\Psi,\\
&\delta{\Chi}=\frac{1}{4}\lambda^{\mu\nu}\bar\sigma_{\mu\nu}\Chi=\frac{1}{4}\lambda^{\mu\nu}\bar\sigma_\mu\sigma_\nu{\Chi}.
\end{split}
\ee
Note that for $\Al=\Oct$, the Lorentz transformations in
(\ref{LTS1}) are not well defined, since they are cubic in octonionic quantities. The choice of association $(\sigma_{[\mu}\sigma_{\nu]})\Psi$ is wrong, since by studying the Fano plane we see that this gives only 31 independent generators when we expect $45=\dim[\SO(1,9)]$. Note that the 14 generators we are missing are the generators of $G_2$, the automorphism group of the octonions. It is fairly straightforward to check that the correct answer is given by $\sigma_{[\mu}(\sigma_{\nu]}\Psi)$; the missing $G_2$ is encoded in the non-associativity of the octonions.

We can think of the Lorentz generators as octonionic operators:
\be
\begin{split}
\hat\sigma_{\mu\nu}&=\frac{1}{2}\Big[\sigma_\mu(\bar\sigma_\nu\cdot)-\sigma_\nu(\bar\sigma_\mu\cdot)\Big],\\
\hat{\bar\sigma}_{\mu\nu}&=\frac{1}{2}\Big[\bar\sigma_\mu(\sigma_\nu\cdot)-\bar\sigma_\nu(\sigma_\mu\cdot)\Big],
\end{split}
\ee
where we adopt the notation that octonionic operators are written with hats. We will also require the octonionic matrices
\be
\begin{split}
\sigma_{\mu\nu}&=\frac{1}{2}(\sigma_\mu\bar\sigma_\nu-\sigma_\nu\bar\sigma_\mu),\\
\bar\sigma_{\mu\nu}&=\frac{1}{2}(\bar\sigma_\mu\sigma_\nu-\bar\sigma_\nu\sigma_\mu),
\end{split}
\ee
which are the generators of $\SL(2,\Oct)/G_2$. It is not a surprise that these simple octonionic matrices do not generate the whole of the $D=10$ Lorentz group, since the space of octonionic $2\times2$ matrices is only 32-dimensional, while $\SO(1,9)$ is 45-dimensional (the 31 independent $\sigma_{\mu\nu}$ are the basis for the space of octonionic $2\times2$ matrices with real trace zero). Since $\R,\C$ and $\Q$ are associative, only in the octonionic case is there a distinction between the action of the sigmas with and without hats.

We conclude that the transformations of the $S_+$ and $S_-$ of $\SL(2,\Al)$ are
\be
\begin{split}\label{LTS2}
\delta\Psi&=\frac{1}{4}\lambda^{\mu\nu}\hat{\sigma}_{\mu\nu}\Psi\hspace{0.3mm}\equiv\frac{1}{4}\lambda^{\mu\nu}\sigma_\mu(\bar\sigma_\nu\Psi),\\
\delta{\Chi}&=\frac{1}{4}\lambda^{\mu\nu}\hat{\bar\sigma}_{\mu\nu}{\Chi}\equiv\frac{1}{4}\lambda^{\mu\nu}\bar\sigma_\mu(\sigma_\nu{\Chi}).
\end{split}
\ee
These transformations allow for a unified treatment of Lorentz transformations for spinors in $D=3,4,6,10$. The fields are always $2\times1$ columns and the Pauli matrices are always $2\times 2$; the only change is the division algebra over which they are defined (and the appearance of brackets in the octonionic case).

One can check that the generators $\frac{1}{2}\hat\sigma_{\mu\nu}$ and $\frac{1}{2}\hat{\bar\sigma}_{\mu\nu}$ satisfy the Lorentz algebra
\be
\begin{split}\label{lorentzalg}
\frac{1}{4}[\hat\sigma^{\mu\nu},\hat\sigma^{\rho\sigma}]&=\frac{1}{2}(\eta^{\sigma\mu}\hat\sigma^{\rho\nu}+
\eta^{\nu\sigma}\hat\sigma^{\mu\rho}-\eta^{\rho\mu}\hat\sigma^{\sigma\nu}-\eta^{\nu\rho}\hat\sigma^{\mu\sigma}),\\
\frac{1}{4}[\hat\sigmabar^{\mu\nu},\hat\sigmabar^{\rho\sigma}]&=\frac{1}{2}(\eta^{\sigma\mu}\hat\sigmabar^{\rho\nu}+
\eta^{\nu\sigma}\hat\sigmabar^{\mu\rho}-\eta^{\rho\mu}\hat\sigmabar^{\sigma\nu}-\eta^{\nu\rho}\hat\sigmabar^{\mu\sigma})
\end{split}
\ee
by acting successively on an arbitrary spinor with the Pauli matrices.

\subsection{Vectors and 2-forms}\label{VECTORS}

Once the spinor and conjugate spinor transformations are known, it is straightforward
to construct the vector representation. As described above, a vector $A$ transforming under $\SO(1,n+1)$ can be parametrised by a $2\times2$ Hermitian matrix using the Pauli basis:
\be
\begin{split}
&A=A^{\mu}\bar\sigma_{\mu}=\begin{pmatrix}
+A^0+A^{n+1} & A^{a+1}e_a^* \\
A^{a+1}e_a & +A^0-A^{n+1} \end{pmatrix},
%\hspace{0.2cm}\text{i.e.}\hspace{0.2cm}A\in{\mathfrak{h}_2(\Al)}
\\
&\bar{A}=A^\mu\sigma_\mu=\begin{pmatrix}
-A^0+A^{n+1} & A^{a+1}e_a^* \\
A^{a+1}e_a & -A^0-A^{n+1} \end{pmatrix}=A-(\Tr A)\id.
\end{split}
\ee
Again, in order to write down an $\mathfrak{sl}(2,\Al)$ transformation we
must generalise the $D=4$ equation
\be
\delta{A}=\frac{1}{4}\lambda^{\mu\nu}\Big(\sigma_{\mu\nu}A-A\bar\sigma_{\mu\nu}\Big).
\ee
Once again this relation is not well defined when $\Al=\Oct$. One might expect the placement of brackets to once again go from right to left. In fact, this bracket placement does give the correct transformation; it can be checked by defining a Hermitian matrix
$\bar{A}=i(\Psi{\Chi}^\dag-\Chi\Psi^\dag)$ and varying using the transformations (\ref{LTS2}). The result confirms:
\be
\delta{A}=\frac{1}{4}\lambda^{\mu\nu}\Big(\hat{\sigma}_{\mu\nu}A-{A}\bar\sigma_{\mu\nu}\Big)
\equiv\frac{1}{4}\lambda^{\mu\nu}\Big(\sigma_\mu(\bar\sigma_\nu{A})-A(\bar\sigma_\mu\sigma_\nu)\Big).
\ee
So we see that the transformation is again a straightforward generalisation of the familiar $D=4$ case.

Using the operators $\hat{\sigma}^{\mu\nu}$ and $\hat{\sigmabar}^{\mu\nu}$ we can also describe spacetime 2-forms such as a field strength $F_{\mu\nu}$:
\be
\begin{split}
&\hat{F}=\frac{1}{2}F_{\mu\nu}\hat\sigma^{\mu\nu},\\
&\hat{\bar{F}}=\frac{1}{2}F_{\mu\nu}\hat{\bar\sigma}^{\mu\nu}.
\end{split}
\ee
These transform under commutation using the Lorentz algbra (\ref{lorentzalg}):
\be
\delta \hat{F}=\frac{1}{4}\lambda^{\mu\nu}[\hat{F},\hat\sigma_{\mu\nu}],~~~~~~\delta\hat{\bar{F}}=\frac{1}{4}\lambda^{\mu\nu}[\hat{\bar{F}},\hat\sigmabar_{\mu\nu}].
\ee

\subsection{Little Groups}\label{LITTLE}

Now that we have a description of the spacetime fields required for super Yang-Mills, as well as their transformations, it will be useful to study their decompositions into the little group. Since the fields we are concerned with are all massless, and we are working in $D=n+2$, the non-trivial little group\footnote{Note, the little group is ISO$(D-2)$, but we neglect the translation generators since they annihilate the physical states leaving only $\SO(D-2)$ with a non-trivial action.} is $\SO(n)$, which has an obvious action on the $n$-dimensional divison algebra $\Al$; the vector, spinor and conjugate spinor representations of $\SO(n)$ each correspond to a single copy of $\Al$. Furthermore, in the language of little group representations we will see that the notion of a triality algebra as defined in (\ref{TRIDEF}) arises naturally in the context of supersymmetric theories.

To truncate $\SO(1,1+n)$ to $\SO(n)$ we can just set $\lambda^{0\mu}=\lambda^{n+1,\mu}=0$. For notational convenience we define the parameters
\be
\theta^{ab}\equiv\lambda^{a+1,b+1}.
\ee
Then, the spinor transforms as
\be
\delta\Psi=\frac{1}{4}\lambda^{\mu\nu}\hat{\sigma}_{\mu\nu}\Psi\hspace{0.2cm}\Rightarrow\hspace{0.2cm}\delta\begin{pmatrix}
\psi \\ 
\chi \end{pmatrix}=\frac{1}{4}\theta^{ab}\begin{pmatrix}
e_a^*(e_b\psi) \\
e_a(e_b^*\chi) \end{pmatrix},
\ee
from which we deduce the transformations of the spinor $\psi$ and conjugate spinor $\chi$ representations of $\SO(n)$:
\be
\begin{split}\label{LGLTS}
\delta\psi&=\frac{1}{4}\theta^{ab}e_a^*(e_b\psi),\\
\delta\chi&=\frac{1}{4}\theta^{ab}e_a(e_b^*\chi).
\end{split}
\ee
Similarly, with $\lambda^{0\mu}=\lambda^{n+1,\mu}=0$, the vector transforms as:
\be
\begin{split}
\delta\begin{pmatrix}
A^0+A^{n+1} & a^* \\
a & A^0-A^{n+1}
\end{pmatrix}&=\frac{1}{4}\theta^{ab}\begin{pmatrix}
0 & e_a^*(e_ba^*)-a^*(e_ae_b^*) \\
e_a(e_b^*a)-a(e_a^*e_b) & 0
\end{pmatrix},
\end{split}
\ee
which tells us the transformation of the vector $a$ of $\SO(n)$:
\be\label{LGLTV}
\delta a = \frac{1}{4}\theta^{ab}\Big(e_a(e_b^*a)-a(e_a^*e_b)\Big).
\ee
The simplicity of these transformations reflects the fact that the structure constants in equation (\ref{STCON}) satisfy
\begin{align}\label{CLIFFORD}
&\Gamma^a\bar\Gamma^b+\Gamma^b\bar\Gamma^a=2\delta^{ab}\id,\nonumber\\
&\bar\Gamma^a\Gamma^b+\bar\Gamma^b\Gamma^a=2\delta^{ab}\id,
\end{align}
so they can be used to define generators of the spinor and conjugate spinor representations of $\SO(n)$
\cite{dundarer1991octonionic}:
\be
\begin{split}
\Sigma^{[ab]}&\equiv\frac{1}{2}\Gamma^{[a}\bar\Gamma^{b]},\\
\bar\Sigma^{[ab]}&\equiv\frac{1}{2}\bar\Gamma^{[a}\Gamma^{b]}.\label{CSGSO8}
\end{split}
\ee
Their components\footnote{We adopt the notation of writing antisymmetric generator labels in square brackets to distinguish them from the generator components.} are given by:
\begin{align}\label{SO8COMPONENTS}
&\Sigma^{[ab]}_{cd}=\delta_{c[a}\delta_{b]d}-\delta_{0[a}C_{b]cd}+\delta_{0[c}C_{d]ab}-\frac{1}{2}Q_{abcd}+4\delta_{0[c}\delta_{d][a}\delta_{b]0},\\
&\bar\Sigma^{[ab]}_{cd}=\delta_{c[a}\delta_{b]d}+\delta_{0[a}C_{b]cd}+\delta_{0[c}C_{d]ab}-\frac{1}{2}Q_{abcd}.
\end{align}
We thus have the interpretation that multiplying a divison algebra element $\psi$ by the basis element $e_a$ has the effect of multiplying $\psi$'s components by the matrix $\bar\Gamma^a$:
\be\label{PSIMULT}
e_a\psi=e_a e_b \psi_b= \Gamma^a_{bc}e_c \psi_b = e_c\bar\Gamma^a_{cb}\psi_b.
\ee
With this in mind the spinor transformations (\ref{LGLTS}) make perfect sense. As for the vector, it is not hard to show (using the identities provided in Section \ref{NDA}) that the components of $a$ transform as
\be
\delta a = \frac{1}{4}\theta^{ab}\Big(e_a(e_b^*a)-a(e_a^*e_b)\Big)=e_a \theta^{ab} a_b,
\ee
so the generators of the vector representation of $\SO(n)$ are:
\begin{align}\label{VGSO8}
J_{[ab]cd}\equiv\delta_{ca}\delta_{bd}-\delta_{cb}\delta_{ad},
\end{align}
as required. We summarise the numbers of components of the $\SO(1,n+1)$ and $\SO(n)$ representations in Table \ref{COUNTING} \cite{freedman2012supergravity,deWit:2002vz}.

\begin{table}[h!]
\begin{center}
\begin{tabular}{c|c|c|c|c}
\hline
\hline
\hspace{0.2cm}$D=n+2$\hspace{0.2cm} & \hspace{0.2cm}Field\hspace{0.2cm} & \hspace{0.2cm}Components\hspace{0.2cm} & Little group field & \hspace{0.2cm}Components\hspace{0.2cm}
\\
\hline
$10=8+2\hspace{0.8mm}$ & $\Psi_{\Oct}$ & $16=2\times8$ & $\psi_{\Oct}$ & $8$ \\
& $\Chi_{\Oct}$ & $16=2\times8$ & $\chi_{\Oct}$ & $8$ \\
& $A_{\Oct}$ & $10$ & $a_{\Oct}$ & $8$ \\ \hline
$6=4+2$ & $\Psi_{\Q}$ & $8=2\times4$ & $\psi_{\Q}$ & $4$ \\
& $\Chi_{\Q}$ & $8=2\times4$ & $\chi_{\Q}$ & $4$ \\
& $A_{\Q}$ & $6$ & $a_{\Q}$ & $4$ \\ \hline
$4=2+2$ & $\Psi_{\C}$ & $4=2\times2$ & $\psi_{\C}$ & $2$ \\
& $\Chi_{\C}$ & $4=2\times2$ & $\chi_{\C}$ & $2$ \\
& $A_{\C}$ & $4$ & $a_{\C}$ & $2$ \\ \hline
$3=1+2$ & $\Psi_{\R}$ & $2=2\times1$ & $\psi_{\R}$ & $1$ \\
& $A_{\R}$ & $3$ & $a_{\R}$ & $1$ \\ \hline\hline
\end{tabular}
\caption{\footnotesize{Spacetime fields over the normed divison algebras}}\label{COUNTING}
\end{center}
\end{table}

Having established the necessary $\SO(n)$ little group transformations, note that we get the spinor and vector representations of $\SO(n-1)$ for free (ignoring the $n=1$ case). The $i$ component of $\Gamma^a$ (and of $-\bar\Gamma^a$) is:
\begin{flalign}
\Gamma^i_{ab}=-\Gamma^i_{ba}=\delta_{0a}\delta_{ib}-\delta_{ia}\delta_{0b}+C_{iab}.
\end{flalign}
These matrices appear when we multiply general basis elements with imaginary basis elements:
\be
e_ie_a=\Gamma^i_{ab}e_b=-\Gamma^i_{ba}e_b.
\ee
It follows directly from (\ref{CLIFFORD}) that these matrices satisfy the
$\SO(n-1)$ Clifford Algebra:
\be\label{CLIFFORD2}
\Gamma^i\Gamma^j+\Gamma^j\Gamma^i=-2\delta^{ij}\id,
\ee
and the corresponding generators are
\begin{flalign}\label{SGSO7}
\Sigma^{[ij]}\equiv\frac{1}{2}\Gamma^{ij}=\frac{1}{2}\Gamma^{[i}\Gamma^{j]}.
\end{flalign}
Their components can easily be seen from (\ref{SO8COMPONENTS}) to be
\be
\Sigma^{[ij]}_{ab}=\delta_{a[i}\delta_{j]b}+\delta_{0[a}C_{b]ij}-\frac{1}{2}Q_{ijab}.
\ee
On setting $\theta^{0i}=0$, the spinor and conjugate spinor transformations (\ref{LGLTS}) coincide, as required, since there is only one spinor representation of $\SO(n-1)$ (for example, in the octonionic case the $\bold{8}_s$ and $\bold{8}_c$ of $\SO(8)$ both become the $\bold{8}$ of SO(7)):
\be
\delta\psi=-\frac{1}{4}\theta^{ij}e_i(e_j\psi).
\ee
For the vector, equation (\ref{LGLTV}) with $\theta^{0i}=0$ transforms only the $n-1$ imaginary parts of $a$, so we conclude that the vector of $\SO(n-1)$ is an imaginary division algebra element, transforming as
\be
\delta a = -\frac{1}{4}\theta^{ij}\Big(e_i(e_ja)-a(e_ie_j)\Big)=e_i \theta^{ij} a_j,
\ee
i.e. under the usual generators $J_{[ij]kl}=\delta_{ki}\delta_{jl}-\delta_{kj}\delta_{il}$.

When working with the little group $\SO(n)$ with division algebras we have the remarkable feature that the vector, spinor and conjugate spinor, as well as the gamma matrices that transform them, are all represented by division algebra elements and their multiplication. We will see in the next section that this formalism is also suited to describing the subgroups that emerge in dimensional reduction, such as $\SO(6)\cong\SU(4)$ in the $D=4$, $\susy=4$ theory.

\section{Symmetries of Super Yang-Mills Theories in $D=n+2$}\label{SYMMETRIES}

In this section, we will dimensionally reduce the little group fields of the octonionic $D=10$ super Yang-Mills theory using the language of division algebras. To motivate this, consider dimensional reduction from $D=10$ to $D=6$, meaning the little group SO(8) becomes $\SO(4)_{ST}\times\SO(4)_I\cong \Sp(1)^2_{ST}\times\Sp(1)^2_I$, where the subscript $ST$ denotes the spacetime little group and the subscript $I$ denotes internal symmetry. The fields in ten dimensions are octonionic while the fields in six dimensions are quaternionic, so we will need to Cayley-Dickson-halve: write an octonion as a pair of quaternions or a kind of `complex quaternion' (not to be confused with a bi-quaternion). Similarly in four dimensions, where the fields are complex, we will view the octonion as a kind of quaternionic complex number. Finally, in three dimensions, the octonion will be viewed as eight real numbers.

Table \ref{TSYG} contains the symmetry groups of the on-shell degrees of freedom for the relevant theories. Each slot corresponds to a pair of division algebras: one to specify $\susy$ and a subalgbra to specify $D$. The algebra representing $D=n+2$ is of course $\Al_n$, while the algebra for $\susy$ is the spacetime algebra of the oxidation endpoint of the theory, $\Al_{n\susy}$, where we use subscripts to denote the dimension of the division algebras. Note that in the table orthogonal groups appear as symmetries in $D=3$, unitary groups appear in $D=4$ and symplectic groups appear in $D=6$. This is of course no coincidence, since $\SO(N)$ is the group of rotations in a real $N$-dimensional
space, $\SU(N)$ is the group of rotations in a complex $N$-dimensional space
and $\Sp(N)$ is the group of rotations in a quaternionic $N$-dimensional
space \cite{Georgi:514148}.

\begin{table}[h!]
\begin{center}
\begin{tabular}{C{1.5cm}|C{3cm}C{3cm}C{3cm}C{3cm}}
\hline\hline
\\
$D\hspace{0.1cm}\backslash\hspace{0.1cm}\mathcal{N}$ & $1$ & $2$ & $4$ &
$8$ \\ \\
\hline \\
$10$ & $\SO(8)_{ST}$ \\ \\
$6$ & $\Sp(1)^2_{ST}\times\Sp(1)_I$ & $\Sp(1)^2_{ST}\times\Sp(1)^2_I$
\\ \\
$4$ & $\Un(1)_{ST}\times\Un(1)_I$ & $\Un(1)_{ST}\times\Un(2)_I$
& $\Un(1)_{ST}\times\SU(4)_I$ \\ \\
$3$ & $1$ & $\SO(2)_I$ & $\SO(4)_I$ & $\SO(7)_I$ \\ \\
\hline\hline
\end{tabular}
\caption{\footnotesize{Spacetime little groups and internal symmetry groups \cite{Strathdee:1986jr,freedman2012supergravity}}}\label{TSYG}
\end{center}
\end{table}

\subsection{$D=10,\hspace{0.2cm}\mathcal{N}=1$}\label{SYMD10}

We will treat each dimension one-by-one, starting from the top. Although the transformations in $D=10$ are special cases of those given above, we include them in this section for completeness. The little group is $\SO(8)$ and the on-shell content of super Yang-Mills is the vector $\bold{8}_v$ and spinor $\bold{8}_s$, each parametrised by a single octonion. We denote these by:
\be
\begin{split}
a_\Oct&=a_ae_a,\\
\psi_{\Oct}&=\psi_ae_a.
\end{split}
\ee
Simple application of (\ref{LGLTV}) and (\ref{LGLTS}) gives their respective $\SO(8)$ transformations. For the spinor:
\be
\begin{split}
\delta\psi_\Oct=\frac{1}{4}\theta^{ab}e_a^*(e_be_c)\psi_c
=\frac{1}{4}\theta^{ab}e_f\gamo{a}{f}{d}\bgamo{b}{d}{c}\psi_c
=\frac{1}{2}\theta^{ab}e_d\Sigma^{[ab]}_{dc}\psi_c,
\end{split}
\ee
while for the vector:
\be
\begin{split}
\delta{a}_\Oct&=\frac{1}{4}\theta^{ab}\Big(e_a(e_b^*e_c)-e_c(e_a^*e_b)\Big)a_c=\frac{1}{4}\theta^{ab}e_f(\bgamo{b}{c}{d}\gamo{a}{d}{f}-\bgamo{a}{b}{d}\gamo{c}{d}{f})a_c\\
&=\frac{1}{4}\theta^{ab}e_f(2\delta_{fa}\delta_{bc}-2\delta_{fb}\delta_{ac})a_c\hspace{0.08cm}=\frac{1}{2}\theta^{ab}e_dJ_{[ab]dc}a_c.
\end{split}
\ee

\subsection{From $D=10,\hspace{0.2cm}\mathcal{N}=1$ to $D=6,\hspace{0.2cm}\mathcal{N}=1,2$}\label{SYMD6}

Taking the fields to be independent of four directions, we arrive at the maximal $D=6$, $\mathcal{N}=2$
theory, which we will formulate over the octonions. The minimal $\mathcal{N}=1$ theory can be obtained by
truncating the octonions to quaternions, i.e. by restricting to a single line of the Fano plane. The spinor and vector decompose as: 
\be
\begin{split}
{\SO(8)}_{ST}&\supset\Sp(1)^2_{ST}\times\Sp(1)^2_I\\
\bold{8}_s&\rightarrow\bold{(2,1;2,1)+(1,2;1,2)},\\
\bold{8}_v&\rightarrow\bold{(2,2;1,1)+(1,1;2,2)},
\end{split}
\ee
i.e. a spinor reduces to
a spinor and a conjugate spinor, while the vector reduces to a vector and four scalars. Each of these four-dimensional irreps will be parameterised by a quaternion, so we need a way to write an octonion as a pair of quaternions. To see how this works, consider a general octonion $x_\Oct=x_ae_a$, $a=0,1,\ldots,7$, and choose a line of the Fano plane, say 457. Then $\Q\cong\Span\{e_0,e_4,e_5,e_7\}$, so we can write:
\be
\begin{split}
x_\Oct&=x_0+x_1e_1+x_2e_2+x_3e_3+x_4e_4+x_5e_5+x_6e_6+x_7e_7\\
&=(x_0+x_4e_4+x_5e_5+x_7e_7)+e_3(x_3+x_6e_4+x_2e_5+x_1e_7),
\end{split}
\ee
where we have chosen $e_3$ to factorise the terms corresponding to the quadrangle $1236$. The octonion $x_\Oct$ now looks like a complex quaternion, with $e_3$ singled out as the imaginary unit separating the two quaternions (of course we could have picked any other of the basis elements and any line of the Fano plane to begin with). The division algebra describing spacetime in $D=6$ is $\Q$, so, to reach this from $\Oct$, we will dimensionally reduce along the directions associated with the $1236$ quadrangle.

We can write this compactly if we define indices
\be
\begin{split}
\hat{a}=0,4,5,7~~~\text{and }~~\hat{i}=4,5,7,\\
\check{a}=1,2,3,6~~~\text{and }~~\check{i}=1,2,6,
\end{split}
\ee
so that `line indices' with hats correspond to spacetime directions and `quadrangle indices' with inverted hats correspond to internal directions. Then we write:
\be
x_\Oct=x_{\ahat} e_{\ahat}+x_{\av} e_{\av}=(x_{\ahat}-e_3x_{\av}\Gamma^3_{\av\ahat})e_{\ahat}.
\ee
Using this, the octonionic vector and spinor can be rearranged to look like:
\be
\begin{split}\label{SIXSPLIT}
a_\Oct&=a_\Q+\phi_{{\Q}^{\complement}}\hspace{0.2cm}=a_{\hat{a}}e_{\hat{a}}+\phi^{\check{a}}e_{\check{a}},\\
\psi_\Oct&=\psi_\Q+e_3\chi_\Q\hspace{-0.02cm}=\psi_{\hat{a}}\eh{a}+e_3(\chi_{\hat{a}}e_{\hat{a}}),
\end{split}
\ee
where $\Q^\complement\cong e_3\Q$ is the complement of the $\Q$ subalgebra spanned by $\{e_0,e_4,e_5,e_7\}$ in $\Oct$. It will sometimes be useful to think of the scalars as a quaternion $\phi_\Q$ and sometimes as an octonion $\phi_{{\Q}^{\complement}}$ living in the particular subspace $\Q^\complement$:
\be
\phi_{{\Q}^{\complement}}=e_3\phi_\Q.
\ee
Note that writing an upper internal index for the scalars in the definition (\ref{SIXSPLIT}) serves just as a reminder that they transform only under internal symmetries (indices are raised and lowered with $\delta_{\ahat\bhat}$ and $\delta_{\av\bv}$). 

Partitioning the Fano plane into a line (plus $e_0$) and quadrangle and studying the multiplication of basis elements, we conclude that multiplying:
\begin{itemize}
\item{two elements on the line returns an element on the line}
\item{two elements on the quadrangle returns an element on the line}
\item{an element from the line and an element from the quadrangle returns an element on the quadrangle}.
\end{itemize}
In terms of the structure constants this translates into:
\be
\begin{split}
&e_{\hat{a}}e_{\hat{b}}=\gamquuu{a}{b}{c}e_{\hat{c}},~~~~~~~~e_{\hat{a}}e_{\check{b}}=\gamqudd{a}{b}{c}e_{\check{c}},\\
&e_{\check{a}}e_{\hat{b}}=\gamqdud{a}{b}{c}e_{\check{c}},~~~~~~~~e_{\check{a}}e_{\check{b}}=\gamqddu{a}{b}{c}e_{\hat{c}}.
\end{split}
\ee
Factorisation of $e_3$ works as follows:
\be
\begin{split}\label{GAMT}
&e_3e_{\hat{a}}=\gamtud{a}{b}e_{\check{b}}~~~~\Rightarrow~~~~e_{\hat{a}}=-e_3\gamtud{a}{b}e_{\check{b}},\\
&e_3e_{\check{a}}=\gamtdu{a}{b}e_{\hat{b}}~~~~\Rightarrow~~~~e_{\check{a}}=-e_3\gamtdu{a}{b}e_{\hat{b}},
\end{split}
\ee 
where the second two relations come from multiplying the first two by $e_3$ on the left (and invoking alternativity). 

Now we know how to write an octonion as a pair of quaternions we can investigate the effect of an $ \Sp(1)^2_{ST}\times \Sp(1)^2_I\subset\SO(8)$ transformation on the $\bold{8}_v,\bold{8}_s$ representations. We restrict to this subgroup by splitting the parameters
\be
\theta^{ab}\rightarrow \theta^{\ahat\bhat},\theta^{\av\bv},\theta^{\ahat\av}
\ee
and setting $\theta^{\ahat\av}=0$. Transforming the spinor with (\ref{LGLTS}) then gives:
\be
\begin{split}
\delta\psi_\Oct&=\frac{1}{4}\theta^{ab}e_a^*(e_b\psi_\Oct)\\
&=\frac{1}{4}\theta^{\hat{a}\hat{b}}e_{\hat{a}}^*(e_{\hat{b}}\psi_\Q)
+\frac{1}{4}\theta^{\hat{a}\hat{b}}e_{\hat{a}}^*(e_{\hat{b}}(e_3\chi_\Q))
+\frac{1}{4}\theta^{\check{a}\check{b}}\ef{a}^*(\ef{b}\psi_\Q)+
\frac{1}{4}\theta^{\check{a}\check{b}}\ef{a}^*(\ef{b}(e_3\chi_\Q)\\
&=\big(\theta^{ST}_\Q\psi_\Q+\psi_\Q\theta^I_\Q\big)+e_3\big(\tilde\theta^{ST}_\Q\chi_\Q+\chi_\Q\tilde\theta^I_\Q\big)
\end{split}
\ee
where each of the $\theta$'s defined is an imaginary quaternion (note that Im($\Q)\cong \mathfrak{sp}(1)$):
\begin{eqnarray}
\theta^{ST}_\Q &\equiv& \frac{1}{4}\theta^{\hat{a}\hat{b}}(e_{\hat{a}}^*e_{\hat{b}})=\frac{1}{2}\Big(+\theta^{0\hat{k}}-\frac{1}{2}\theta^{\hat{i}\hat{j}}\varepsilon_{\hat{i}\hat{j}\hat{k}}\Big)\eh{k},\\ 
\tilde\theta^{ST}_\Q &\equiv& \frac{1}{4}\theta^{\hat{a}\hat{b}}(e_{\hat{a}}e_{\hat{b}}^*)=\frac{1}{2}\Big(-\theta^{0\hat{k}}-\frac{1}{2}\theta^{\hat{i}\hat{j}}\varepsilon_{\hat{i}\hat{j}\hat{k}}\Big)\eh{k},\\ 
\theta^I_\Q&\equiv&\frac{1}{4}\theta^{\check{a}\check{b}}(\ef{a}^*\ef{b})=\frac{1}{2}\Big(-\theta^{3\check{i}}C_{3\check{i}\hat{k}}-\frac{1}{2}\theta^{\check{i}\check{j}}C_{\check{i}\check{j}\hat{k}}\Big)\eh{k},\\
\tilde\theta^I_\Q&\equiv&\frac{1}{4}\theta^{\check{a}\check{b}}(\ef{a}^*\ef{b})+\theta^{3\check{i}}(e_3\ef{i})=\frac{1}{2}\Big(+\theta^{3\check{i}}C_{3\check{i}\hat{k}}-\frac{1}{2}\theta^{\check{i}\check{j}}C_{\check{i}\check{j}\hat{k}}\Big)\eh{k}.
\end{eqnarray}
We conclude that $\psi$ and $\chi$ do indeed transform as the $\bold{(2,1;2,1)}$ and $\bold{(1,2;1,2)}$ of $\Sp(1)^4$, respectively:
\be
\begin{split}
\delta\psi_\Q=\theta^{ST}_\Q\psi_\Q+\psi_\Q\theta^I_\Q,\\
\delta\chi_\Q=\tilde\theta^{ST}_\Q\chi_\Q+\chi_\Q\tilde\theta^I_\Q.
\end{split}
\ee
Similarly, transforming the vector with $\theta^{\ahat\av}=0$ gives:
\be
\begin{split}
\delta{a}_\Oct&=\frac{1}{4}\theta^{ab}\Big(e_a(e_b^*a_\Oct)-a_\Oct(e_a^*e_b)\Big)\\
&=\frac{1}{4}\theta^{\hat{a}\hat{b}}\Big(\eh{a}(\eh{b}^*a_\Q)-a_\Q(\eh{a}^*\eh{b})\Big)
+\frac{1}{4}\theta^{\hat{a}\hat{b}}\Big(\eh{a}(\eh{b}^*\phi_{{\Q}^{\complement}})-\phi_{{\Q}^{\complement}}(\eh{a}^*\eh{b})\Big)\\
&\hspace{0.4cm}+\frac{1}{4}\theta^{\check{a}\check{b}}\Big(\ef{a}(\ef{b}^*a_\Q)-a_\Q(\ef{a}^*\ef{b})\Big)
+\frac{1}{4}\theta^{\check{a}\check{b}}\Big(\ef{a}(\ef{b}^*\phi_{{\Q}^{\complement}})-\phi_{{\Q}^{\complement}}(\ef{a}^*\ef{b})\Big)\\
&=\Big(\tilde\theta^{ST}_\Q{a}_\Q-a_\Q\theta^{ST}_\Q\Big)+e_3\Big(\tilde\theta^I_\Q\phi_\Q-\phi_\Q\theta^I_\Q\Big),
\end{split}
\ee
so the vector and scalars transform as the $\bold{(2,2;1,1)}$ and $\bold{(1,1;2,2)}$ of $\Sp(1)^4$:
\be
\begin{split}
\delta{a_\Q}&=\tilde\theta^{ST}_\Q{a}_\Q-a_\Q\theta^{ST}_\Q,\\
\delta\phi_{\Q}&=\tilde\theta^I_\Q\phi_\Q-\phi_\Q\theta^I_\Q.
\end{split}
\ee
It is a straightforward exercise to show that these correspond to the correct
conventional transformations:
\be
\begin{split}
\delta{a_{\hat{c}}}&=\frac{1}{2}\theta^{\hat{a}\hat{b}}J_{[\hat{a}\hat{b}]\hat{c}\hat{d}}a_{\hat{d}},\\
\delta\phi^{\check{c}}&=\frac{1}{2}\theta^{\check{a}\check{b}}J_{[\check{a}\check{b}]\check{c}\check{d}}\phi^{\check{d}}.
\end{split}
\ee
We have now established that the $D=6$ , $\mathcal{N}=2$ theory can be
formulated over $\Oct$. To obtain the $\mathcal{N}=1$ theory we just discard $\chi_\Q$ and $\phi_{{\Q}^{\complement}}$, essentially truncating fields with quadrangle indices. This leaves us a theory formulated over $\Q$, as required. Note that $\psi_\Q$ still transforms under $\theta^I_\Q$, so there is still an internal symmetry of $\Sp(1)$ in the $\mathcal{N}=1$ theory. Of course, the $\susy=1$ theories in $D=3,4,6,10$ can be obtained from one another simply by switching division algebras, but viewing them as truncations of the maximal theory provides a quick way to find their internal symmetries.

\subsection{From $D=10,\hspace{0.2cm}\mathcal{N}=1$ to $D=4,\hspace{0.2cm}\mathcal{N}=1,2,4$}\label{SYMD4}

Next, dropping dependence on six of the ten dimensions yields the
maximal theory $D=4$, $\mathcal{N}=4$. We will formulate the on-shell degrees of freedom of the theory over the octonions. Truncating to a quaternionic subalgebra will result in the $\mathcal{N}=2$ theory and
further truncation to a complex subalgebra will yield the $\mathcal{N}=1$ theory. Now the spacetime little group is $\SO(2)$ and the internal symmetry is $\SO(6)$; the fields decompose as:
\be
\begin{split}
{\SO(8)}_{ST}&\supset\SO(2)_{ST}\times\SO(6)_{I}\cong\Un(1)_{ST}\times\SU(4)_{I}\\
\bold{8}_{s}&\rightarrow\bold{4}_{1/2}+\bold{\bar{4}}_{-1/2}\\
\bold{8}_{v}&\rightarrow\bold{6}_{0}+\bold{1}_{1}+\bold{1}_{-1}.
\end{split}
\ee
The $D=10$ vector becomes a $D=4$ vector and six scalars, while we get four $D=4$ fermions transforming as the fundamental of $\SU(4)$. Since the division algebra associated with $D=4$ is $\C$, we will need to write our octonions as `quaternions of complex numbers'. In practice, this is no different from writing a `complex quaternion'; the difference is simply the way we transform the resulting fields. We now view $e_3$ as the complex unit corresponding to the complex spacetime (we made this particular choice so that we could use the identities of the previous subsection) and define the indices
\be
\begin{split}
\underline{a}&=0,3,\\
\overline{i}&=1,2,4,5,6,7,
\end{split}
\ee
so that those with an under-line correspond to spacetime (of course there is only one transformation parameter $\theta^{03}$) and those with an over-line are internal. We can then write the octonionic fields as
\be
\begin{split}
a_\Oct &=a_\C+\phi_{{\C}^{\complement}}=a_{\underline{a}}\ed{a}+\phi^{\overline{i}}\eu{i},\\
\psi_\Oct&=\psi_\C+\psi_\C'e_4+\psi_\C''e_5+\psi_\C'''e_7=\psi_\C^{\hat{a}}\eh{a}=\psi_{\underline{a}}^{\hat{a}}\ed{a}\eh{a}.
\end{split}
\ee
Multiplying the spinor $\psi_\Oct=\psi_\C^{\hat{a}}\eh{a}$ by $e_{\ibar}$ has the following effect:
\be
\begin{split}\label{UPSILON}
e_{\ibar}\left(\psi_\C^{\hat{a}}\eh{a}\right)&=\psi_\C^{\hat{a}*}(e_{\ibar}\eh{a})=\psi_\C^{\hat{a}*}(\Gamma^{\ibar}_{\ahat\bhat}e_{\bhat}+\Gamma^{\ibar}_{\ahat\bv}e_{\bv})=\psi_\C^{\hat{a}*}(\Gamma^{\ibar}_{\ahat\bhat}-e_3\Gamma^{\ibar}_{\ahat\bv}\Gamma^{3}_{\bv\bhat})e_{\bhat},
\end{split}
\ee
so its complex components $\psi_\C^{\ahat}$ get complex-conjugated and multiplied by the matrix
\be
\Upsilon_{\ahat\bhat}^{\ibar}=-\Upsilon_{\bhat\ahat}^{\ibar}\equiv \Gamma^{\ibar}_{\ahat\bhat}-e_3\Gamma^{\ibar}_{\ahat\bv}\Gamma^{3}_{\bv\bhat}.
\ee
The matrices $\Upsilon^{\ibar}$ and $\overline\Upsilon{}^{\ibar}\equiv\Upsilon^{\ibar*}$ satisfy the relations
\be
\begin{split}
\Upsilon^{\ibar}\overline\Upsilon{}^{\jbar}+\Upsilon^{\ibar}\overline\Upsilon{}^{\jbar}&=-2\delta^{\ibar\jbar}\id,\\
\overline\Upsilon{}^{\ibar}\Upsilon^{\jbar}+\overline\Upsilon{}^{\ibar}\Upsilon^{\jbar}&=-2\delta^{\ibar\jbar}\id,
\end{split}
\ee
and so can be used to form antihermitian, traceless $4\times4$ generators of $\SU(4)$ for the $\bold{4}$ and $\bar{\bold{4}}$ representations:
\be
\begin{split}
T^{[\bar{i}\bar{j}]}_{\hat{a}\hat{b}}=\frac{1}{2}\Upsilon^{[\ibar}_{\ahat\chat}\overline\Upsilon{}^{\jbar]}_{\chat\bhat}=
\Sigma^{[\bar{i}\bar{j}]}_{\hat{a}\hat{b}}+e_3\gamtud{a}{c}\Sigma^{[\bar{i}\bar{j}]}_{\check{c}\hat{b}},\\
\bar{T}^{[\bar{i}\bar{j}]}_{\hat{a}\hat{b}}=\frac{1}{2}\overline\Upsilon{}^{[\ibar}_{\ahat\chat}\Upsilon^{\jbar]}_{\chat\bhat}=
\Sigma^{[\bar{i}\bar{j}]}_{\hat{a}\hat{b}}-e_3\gamtud{a}{c}\Sigma^{[\bar{i}\bar{j}]}_{\check{c}\hat{b}}.
\end{split}
\ee
Note that $\Gamma^3_{ab}$, satisfying $(\Gamma^3)^2=-\id$ plays the role of a complex structure on $\Oct\cong\R^8$. This means that any real $8\times 8$ matrix that commutes with $\Gamma^3$, such as $\Sigma^{[\ibar\jbar]}_{ab}$, can be written as a complex $4\times4$ matrix, like $T^{[\ibar\jbar]}_{\ahat\bhat}$ above. These generators arise when we dimensionally reduce, since we split the parameters $\theta^{ab}$ as follows
\be
\theta^{ab}\rightarrow \theta^{\overline{a}\overline{b}},\theta^{\underline{a}\underline{b}},\theta^{\overline{a}\underline{a}},
\ee
and restrict to the subgroup $\SU(4)\times\Un(1)\subset\SO(8)$ by setting $\theta^{\overline{i}\underline{j}}=0$.
The transformation of the spinor is then:
\be
\begin{split}
\delta\psi_\Oct&=\frac{1}{4}\theta^{ab}e_a^*(e_b\psi_\Oct)\\
&=\frac{1}{2}\theta^{03}e_3(\psi_\C^{\hat{a}}\eh{a})+\frac{1}{4}\theta^{\bar{i}\bar{j}}\eu{i}^*(\eu{j}(\psi_\C^{\hat{a}}\eh{a}))\\
&=\frac{1}{2}\theta^{03}(e_3\psi_\C^{\hat{a}})\eh{a}-\frac{1}{2}\theta^{\bar{i}\bar{j}}\left(T^{[\bar{i}\bar{j}]}_{\hat{a}\hat{b}}\psi_\C^{\hat{b}}\right)\eh{a},
\end{split}
\ee
where we used equation (\ref{UPSILON}) twice. We deduce that the four complex spinors $\psi_\C^{\ahat}$ do indeed transform as the $\bold{4}_{1/2}+\bold{\bar{4}}_{-1/2}$:
\begin{flalign}
\delta\psi_\C^{\hat{a}}=\frac{1}{2}\theta^{03}e_3\psi_\C^{\hat{a}}-\frac{1}{2}\theta^{\bar{i}\bar{j}}T^{[\bar{i}\bar{j}]}_{\hat{a}\hat{b}}\psi_\C^{\hat{b}}.
\end{flalign}
Interestingly, we can also view $\SU(4)\times\Un(1)$ as the subgroup of $\SO(8)$ consisting of matrices that commute with the complex structure $\Gamma^3$, that is, those transformations that treat the 8 real components of an $\SO(8)$ spinor like 4 complex components. The transformation of the vector with $\theta^{\overline{i}\underline{j}}=0$ is:
\be
\begin{split}
\delta{a}_\Oct&=\frac{1}{4}\theta^{ab}\Big(e_a(e_b^*a_\Oct)-a_\Oct(e_a^*e_b)\Big)\\
&=\frac{1}{2}\theta^{03}\Big(-e_3a_\C-a_\C{e_3}\Big)
+\frac{1}{2}\theta^{03}\Big(-e_3\phi_{{\C}^{\complement}}-\phi_{{\C}^{\complement}}{e_3}\Big)\\
&\hspace{0.4cm}+\frac{1}{4}\theta^{\bar{i}\bar{j}}\Big(\eu{i}(\eu{j}^*a_\C)-a_\C(\eu{i}^*\eu{j})\Big)
+\frac{1}{4}\theta^{\bar{i}\bar{j}}\Big(\eu{i}(\eu{j}^*\phi_{{\C}^{\complement}})-\phi_{{\C}^{\complement}}(\eu{i}^*\eu{j})\Big)\\
&=-\theta^{03}e_3a_\C+\frac{1}{2}\theta^{\bar{i}\bar{j}}\eu{l}J_{[\bar{i}\bar{j}]\bar{l}\bar{k}}\phi^{\bar{k}},
\end{split}
\ee
so the $D=4$ vector and scalars transform as the $\bold{1}_{1}+\bold{1}_{-1}$ and $\bold{6}_{0}$ of $\SU(4)\times\Un(1)$, respectively:
\be
\begin{split}
\delta{a_\C}&=-\theta^{03}e_3a_\C,\\
\delta\phi^{\bar{l}}&=\frac{1}{2}\theta^{\bar{i}\bar{j}}J_{[\bar{i}\bar{j}]\bar{l}\bar{k}}\phi^{\bar{k}},
\end{split}
\ee
as required.
The above calculations demonstrate that the $D=4$, $\mathcal{N}=4$ theory can be formulated
over $\Oct$ with generators appearing directly from the octonionic multiplication
rule. To obtain the $\mathcal{N}=2$ theory one simply truncates to a single line of the Fano plane, so that the $\mathcal{N}=2$ theory is written
over $\Q$. This translates into discarding two fermions and the four scalars. The internal symmetry of the resulting $\susy=2$ theory is then the subgroup of $\SU(4)$ that preserves the quaternionic subalgebra and commutes with the complex structure. This gives the R-symmetry $\Un(2)$.

The $\mathcal{N}=1$ theory can then be obtained by further discarding a spinor and the remaining scalars, corresponding to truncating $\Q$ to $\C$. The internal symmetry is U(1), since this theory contains just a single complex spinor.

\subsection{From $D=10,\hspace{0.2cm}\mathcal{N}=1$ to
$D=3,\hspace{0.2cm}\mathcal{N}=1,2,4,8$}\label{SYMD3}

Finally, we will dimensionally reduce to $D=3$, obtaining the $\susy=8$ maximal theory over the octonions. We can then truncate to the $\susy=4,2,1$ theories by replacing octonions by quaternions, complex numbers or real numbers, respectively. This time the relevant decomposition is simply:
\be
\begin{split}
{\SO(8)}_{ST}&\supset\SO(7)_{I}\\
\bold{8}_{s}&\rightarrow\bold{8}\\
\bold{8}_{v}&\rightarrow\bold{1+7}.
\end{split}
\ee
The spacetime little group here is trivial (or SO(1)), so the vector, fermions and scalars each contain only a single on-shell degree of freedom. We split the parameters as
\be
\theta^{ab}\rightarrow\theta^{0i},\theta^{ij}
\ee
and set $\theta^{0i}=0$. This is just the $n=8$ case of the discussion of $\SO(n-1)$ in Section \ref{LITTLE}. We write the fields then as
\be
\begin{split}
a_\Oct&=a_\R+\phi_{{\R}^{\complement}}=a_0+\phi^ie_i,\\
\psi_\Oct&=\psi_\R^ae_a,
\end{split}
\ee
and we find that they transform as:
\be
\begin{split}
\delta\psi_\Oct&=\frac{1}{4}\theta^{ij}e_i^*(e_je_a)\psi_\R^a\\
&=-\frac{1}{2}\theta^{ij}e_b\Sigma^{[ij]}_{ba}\psi_\R^a
\end{split}
\ee
and
\be
\begin{split}
\delta a_\Oct&=\frac{1}{4}\theta^{ij}\Big(e_i(e_j^*a_\R)-a_\R(e_i^*e_j)\Big)+\frac{1}{4}\theta^{ij}\Big(e_i(e_j^*\phi_{{\R}^{\complement}})-\phi_{{\R}^{\complement}}(e_i^*e_j)\Big)\\
&=\frac{1}{2}\theta^{ij}e_lJ_{[ij]lk}\phi^k.
\end{split}
\ee
We conclude that the fermions and scalars transform as the $\bold{8}$ and $\bold{7}$ of SO(7), as required:
\be
\begin{split}
\delta\psi_\R^a&=-\frac{1}{2}\theta^{ij}\Sigma^{[ij]}_{ab}\psi_\R^b,\\
\delta\phi^l &=\frac{1}{2}\theta^{ij}J_{[ij]lk}\phi^k.
\end{split}
\ee
The R-symmetry of the $D=3$ theory \cite{Strathdee:1986jr} (that is, the group of automorphisms of the supersymmetry algebra) is actually SO(8), but we can only see this if we dualise the vector to a scalar. This can only be carried out in the free Yang-Mills theory with coupling constant $g=0$ (see Section \ref{LAGRAS}). 

To obtain the $\mathcal{N}=4$ theory over $\Q$, once again, we simply truncate field content corresponding to a quadrangle of the Fano plane, leaving us with a quaternion of spinors and an imaginary quaternion of scalars. This theory has internal symmetry $\Sp(1)\times\Sp(1)$. The $\mathcal{N}=2$ theory over $\C$ can then be obtained by further discarding the content associated with two of the imaginary elements, i.e. truncating two scalars and two spinors. The internal symmetry is then $\Un(1)$. Finally, truncating the remaining scalar and the spinor associated with
the last imaginary element results in the $\mathcal{N}=1$ theory formulated
over $\R$, with no internal symmetry.
\\
\begin{table}[h!]
\begin{center}
\begin{tabular}{c|c|c|c}
\hline
\hline
\hspace{0.2cm} \hspace{0.2cm} &\hspace{0.4cm}R-symmetry\hspace{0.4cm}
&\hspace{0.1cm} $g\neq0$ Lagrangian\hspace{0.1cm} &\hspace{0.1cm} $g=0$ Lagrangian\hspace{0.1cm}
\\
\hline
$\susy=1$ & $-$ & $ -$ & $-$ \\
$\susy=2$ & U(1) & $\Un(1)$ & $\Un(1)^2$ \\
$\susy=4$ & $\Sp(1)\times\Sp(1)$ & $\Sp(1)\times\Sp(1)$ & $\Sp(1)^3$ \\
$\susy=8$ & SO(8) & $\SO(7)$ & $\SO(8)~$ \\
\hline
\hline
\end{tabular}
\end{center}
\caption{\footnotesize{Symmetries in $D=3$ SYM theories. The symmetries of the $g=0$ Lagrangian are the triality algebras of $\R,\C,\Q,\Oct$, while the symmetries of the $g\neq 0$ Lagrangian are known in the literature as `intermediate algebras' (these are just the subgroups of the triality algebras such that $A(1)=0$ in equation (\ref{TRIDEF}) \cite{Barton:2003,westbury2006sextonions}).}}\label{D=3}
\end{table}
\\
For the cases $\susy=1,2,4$, the groups mentioned in the preceding paragraph agree with the R-symmetry $\SO(\susy)$ expected from a $D=3$ theory with $\susy$ supersymmetries \cite{Strathdee:1986jr}. However, if we set the coupling constant $g=0$ then we can dualise the vector to a scalar and the symmetry is enlarged; $\susy=2$ has $\Un(1)^2$ and $\susy=4$ has $\Sp(1)^3$ (the internal symmetries match the overall symmetries of the physical degrees of freedom of the $D=4,6$ theories, from which they are obtained by dimensional reduction). This is clarified in Table \ref{D=3}.

\subsection{Triality Algebras}

We have demonstrated that minimal SYM in $D=n+2$ can be formulated over the division algebra $\Al_n$, and that doubling the amount of supersymmetry has the effect of Cayley-Dickson-doubling the algebra, with the process terminating at theories with maximal supersymmetry written over the octonions. We can write this somewhat schematically as
\be
\Al_n\Al_\susy\sim\Al_{n\susy},
\ee
with every case given in Table \ref{ALGEBRAS}; the fields are naturally valued in $\Al_n$ and we package $\susy$ of them into an $\Al_{n\susy}$-valued object. 

Looking back at Table \ref{TSYG}, we see that the total (spacetime plus internal) symmetry of the $\susy=1$ theories in $D=n+2$ is given by the triality algebra $\mathfrak{tri}(\Al_n)$ as defined in (\ref{TRIDEF}). To understand this, one must consider the following. When we formulate our vector $a$, spinor $\psi$ and conjugate spinor $\chi$ representations of $\SO(n)$  in terms of division algebras, the relationships between the three may be expressed without gamma matrices. For example, we can define a vector $a$ by
\be\label{tri1}
a=\chi\psi^*= \chi^b\psi^c\bar\Gamma^a_{bc} e_a,
\ee
or we could define a spinor by
\be\label{tri2}
\psi=a^*\chi =a^a\chi^c {\Gamma}^a_{bc} e_b,
\ee
or a conjugate spinor by
\be\label{tri3}
\chi = a\psi = a^a\psi^c\bar\Gamma^a_{bc} e_b.
\ee
Let us choose, say, equation (\ref{tri1}) and consider acting on $a,\psi$ and $\chi$ with three (\emph{a priori} unrelated) $\SO(n)$ transformations,
\be
\delta a = A(a),~~~~\delta \chi=B(\chi),~~~~\delta\psi^*=C(\psi^*),
\ee
while demanding that the left- and right-hand sides transform in the same way. This brings us precisely to the definition of a triality algebra (\ref{TRIDEF}), and we are led to the conclusion that the largest group of transformations that preserves  (\ref{tri1}) is $\mathfrak{tri}(\Al_n)$. However, equations such as (\ref{tri1}) are exactly of the form used in $\susy=1$ supersymmetry transformations, so it is only natural that the overall symmetry of these theories is given by the triality algebras.

Since any SYM theory can be obtained by dimensional reduction of one of the $\susy=1$ theories, we can extract from this a general formula for the total symmetry of the on-shell degrees of freedom for any SYM theory. A subgroup of the triality algebra corresponds to spacetime symmetry and it is this subgroup that is restricted when we dimensionally reduce. The on-shell degrees of freedom of the vector are always just an element of $\Al_n$, written as a subalgebra of $\Al_{n\susy}$, so we can understand dimensional reduction as the condition that this subalgebra be preserved by one element of the triality triple:
\be\label{TRITILDE}
\widetilde{\mathfrak{tri}{}}(\Al_{n\susy},\Al_{n})\equiv\big\{(A,B,C)\in3\hspace{0.5mm}\mathfrak{so}(n\susy)|A(xy)=B(x)y+xC(y) \text{ and } A(\Al_n\subseteq \Al_{n\susy})=\Al_n\big\}, \hspace{0.4cm}x,y\in\Al_{n\susy},
\ee
where $3\hspace{0.5mm}\mathfrak{so}(n\susy)=\mathfrak{so}(n\susy)\oplus\mathfrak{so}(n\susy)\oplus\mathfrak{so}(n\susy)$. This definition provides a concise summary of Table \ref{TSYG}, giving the symmetry of any Yang-Mills theory in $D=n+2$. In $D=3$, after dualising the vector to a scalar, the full triality algebras appear as symmetries, as in \cite{Borsten:2013bp}; they are simply inherited from the $\susy=1$ theories in the higher dimensions. Note that the definition (\ref{TRITILDE}) would give the same Lie algebras if we had picked $B$ or $C$ to preserve the $\Al_n$ subalgebra, due to the principles of triality outlined in Appendix \ref{APPA}.

\begin{center}
\begin{table}[h!]
\begin{tabular}{C{1.5cm}|C{2.5cm}C{2.5cm}C{2.5cm}C{2.5cm}}
\hline\hline
\\
$D\hspace{0.1cm}\backslash\hspace{0.1cm}\mathcal{N}$
& $1$ & $2$ & $4$ & $8$ \\ \\
\hline \\
$10$ & $\Oct\R\sim\Oct$ \\ \\
$6$ & $\Q\R\sim\Q$ & $\Q\C\sim\Oct$
\\ \\
$4$ & $\C\R\sim\C$ & $\C\C\sim\Q$
& $\C\Q\sim\Oct$ \\ \\
$3$ & $\R\R\sim\R$ & $\R\C\sim\C$ & $\R\Q\sim\Q$ & $\R\Oct\sim\Oct$
\\ \\
\hline\hline
\end{tabular}
\caption{\footnotesize{A summary of the division algebras used in $D=n+2$ SYM with $\susy$ supersymmetries: $\Al_n\Al_\susy\sim\Al_{n\susy}$.}}\label{ALGEBRAS}
\end{table}
\end{center}

\section{Lagrangians of Super Yang-Mills Theories in $D=n+2$}\label{LAGRAS}

\subsection{$\mathcal{N}=1$ Theories}\label{LAGN1}

In this section we will focus on the action and supersymmetry transformations of Yang-Mills in $D=n+2$. To write a spinor kinetic term in $D=n+2$ we require a real, Lorentz-scalar spinor bilinear. Such a product is given by \cite{Schray:1994ur,Baez:2009xt}
\be
\text{Re}(i\Psi^\dagger\Chi)=\frac{i}{2}(\Psi^\dagger\Chi-\Chi^\dagger\Psi),
\ee
where $\Psi$ belongs to the spinor $S_+$ representation and $\Chi$ to the conjugate spinor representation $S_-$. It is simple to verify that this is Lorentz-invariant using the transformations (\ref{LTS2}). Note that the components $\Psi^a$ here are anti-commuting, so we are dealing with the division algebras defined over the Grassmanns rather than the reals. The appearance of the complex unit $i$ in the above equation is independent of the division algebra $\Al_n$. This is only an artifact of our spinor conventions.

The spinor Lagrangian is then the product of the $S_+$ spinor $\Psi$ and the $S_-$ conjugate spinor $\sigmabar^\mu\partial_\mu\Psi$:
\be
-\text{Re}(i\Psi^{\dagger}\sigmabar^\mu\partial_\mu\Psi)=-\frac{i}{2}\Psi^{\dagger}(\sigmabar^\mu\partial_\mu\Psi)-\frac{i}{2}(\Psi^\dagger\sigmabar^\mu)\partial_\mu\Psi + \text{ total derivative},
\ee
where we omit the association brackets on the left-hand side since the associator is pure-imaginary:
\be\label{RE}
\text{Re}((ab)c)=\text{Re}(a(bc))\equiv\text{Re}(abc).
\ee
The overall sign ensures we agree with the usual $D=4$ expression $-i\Psi^{\dagger}\sigmabar^\mu\partial_\mu\Psi$. The action for $(n+2)$-dimensional $\mathcal{N}=1$ SYM with gauge group $G$ over the division algebra $\Al_n$ is then given by
\begin{equation}\label{10-d action}
S=\int d^{n+2}x\left(-\frac{1}{4}F_{\mu\nu}^AF^{A\mu\nu}-\text{Re}(i\Psi^{\dagger A}\sigmabar^\mu D_\mu\Psi^A)\right),~~~~~\Psi\in\Al_n^2,
\end{equation}
where $A=0,\dots,\dim[G]$ and the covariant derivative and field strength are given by the usual expressions
\be
\begin{split}
D_\mu\Psi^A&=\partial_\mu\Psi^A+gf_{BC}{}^A A_\mu^B\Psi^C,\\
F^A_{\mu\nu}&=\partial_\mu A^A_\nu-\partial_\nu A_\mu^A+gf_{BC}{}^AA^B_\mu A^C_\nu.
\end{split}
\ee
The supersymmetry transformations are
\be\label{SUSYTRANS}
\delta A_\mu^A=\text{Re}(i\Psi^{A\dagger}\sigmabar_\mu \epsilon),\hspace{0.5cm}
\delta\Psi^A=\frac{1}{2}\hat{F}^A\epsilon.
\ee
Another strength of using division algebraic spinors is that we may write the vector's transformation without $\sigma$ matrices simply by taking the outer product:
\be
\delta \bar{A}^A = \delta A^A_{\mu}\sigma^\mu =i(\Psi^A\epsilon^\dagger-\epsilon\Psi^{\dagger A}).
\ee
To write this in terms of $A^A=A_\mu\sigmabar^\mu$ we just reverse the trace:
\begin{equation}\label{variation no sigma}
\begin{split}
\delta A^A=\delta A_\mu^A\sigmabar^\mu&=i(\Psi^A\epsilon^\dagger-\epsilon\Psi^{\dagger A})-(\text{trace})\\
&=i(\Psi^A\epsilon^\dagger-\epsilon\Psi^{\dagger A})+i(\epsilon^\dagger\Psi^A-\Psi^{\dagger A}\epsilon)\id,
\end{split}
\end{equation}
where the trace term is calculated using the cyclicity of the real trace, while taking into account the Grassmann nature of the spinors:
\be
\tr i(\Psi^A\epsilon^\dagger-\epsilon\Psi^{\dagger A})=\text{Re}\tr i(\Psi^A\epsilon^\dagger-\epsilon\Psi^{\dagger A}) =
-\text{Re}\tr i(\epsilon^\dagger\Psi^A-\Psi^{\dagger A}\epsilon)= - i(\epsilon^\dagger\Psi^A-\Psi^{\dagger A}\epsilon).
\ee

\subsection{The Master Lagrangian}\label{LAGM}

We can now dimensionally reduce this Lagrangian using the techniques described in the previous section to obtain a `master Lagrangian', whose input is the division algebra $\Al_{n\susy}$ and its subalgebra $\Al_n$, and whose output is the Yang-Mills theory in $D=n+2$ with $\susy$ supersymmetries. The vector decomposes as follows:
\be
\begin{split}
A_{\Al_{n\susy}} =\begin{pmatrix}
A^0+A^{n\susy+1} & a_{\Al_{n\susy}}^* \\
a_{\Al_{n\susy}} & A^0-A^{n\susy+1} \end{pmatrix}&=
\begin{pmatrix}
A^0+A^{n\susy+1} & a_{\Al_{n}}^* \\
a_{\Al_{n}} & A^0-A^{n\susy+1} \end{pmatrix}+
\begin{pmatrix} 0 & \phi_{\Al_{n}^\complement}^* \\
\phi_{\Al_{n}^\complement} & 0 \end{pmatrix}\\&=A_{\Al_n}+\phi_{\Al_{n}^\complement}\eps,
\end{split}
\ee
where $\phi_{\Al_{n}^\complement} \in\Al_n^\complement=\Al_{n\susy}\backslash\Al_n$. We leave the spinor $\Psi\in\Al_{n\susy}^2$ just as it is, understanding that we have $\susy$ spinors in $D=n+2$ dimensions, each valued in $\Al_{n}^2$. Dropping the subscript on $\phi$, the resulting action is
\be
\begin{split}\label{MASTER}
S\left(\Al_n,\Al_{n\susy}\right)=\int d^{n+2}x&\left(-\frac{1}{4}F^A_{\mu\nu}F^{A\mu\nu}-\frac{1}{2}D_\mu\phi^{A*}D^\mu\phi^A-\text{Re}(i\Psi^{\dagger A}\sigmabar^\mu D_\mu\Psi^A) \right. \\ 
&-gf_{BC}{}^A\text{Re}\left(i{\Psi}^{\dagger A}\varepsilon\phi^B\Psi^C\right) \\
 & \left.-\frac{1}{16}g^2f_{BC}{}^Af_{DE}{}^A(\phi^{B*}\phi^D+\phi^{D*}\phi^B)(\phi^{C*}\phi^E+\phi^{E*}\phi^C)  \right),
\end{split}
\ee
where $\{\sigmabar^\mu\}$ are a basis for $\Al_n$-valued Hermitian matrices. The supersymmetry transformations are
\be
\begin{split}
\delta \bar{A}^A& =i(\Psi^A\epsilon^\dagger-\epsilon\Psi^{\dagger A})_{\Al_n},\\
\delta\phi^A&=- \frac{i}{2}\tr\left(\eps(\Psi^A\epsilon^\dagger-\epsilon\Psi^{\dagger A})_{\Al_n^\complement}\right) ,\\
\delta\Psi^A&=\frac{1}{2}\hat{F}^A\epsilon+\frac{1}{2}\sigma^\mu\eps(D_\mu\phi^A\epsilon)+\frac{1}{4}f_{BC}{}^A\phi^C(\phi^B\epsilon),
\end{split}
\ee
where the subscripts $\Al_n$ and $\Al_n^\complement$ refer to the respective projections onto these subspaces.

To obtain the conventional actions, one can always multiply out the division algebra basis elements as appropriate to the theory of interest. For example, for $D=4$, $\susy=4$ we have $\Al_{n\susy}=\Al_8=\Oct$ and $\Al_n=\Al_2=\C$, so the fermions look like `complex quaternions'. Multiplying out the quaternionic basis elements $e_{\ahat}$ returns the conventional action, in terms of four complex fermions $\Psi^{\ahat}$ and six real scalars $\phi^{\ibar}$:
\be
\begin{split}\label{N=4}
S\left(\C,\Oct\right)=\int d^{4}x
&\left(-\frac{1}{4}F^A_{\mu\nu}F^{A\mu\nu}-\frac{1}{2}D_\mu\phi^{A\ibar}D^\mu\phi^{A\ibar}-\text{Re}(i\Psi^{\dagger A\ahat}\sigmabar^\mu D_\mu\Psi^{A\ahat})\right.\\ 
&  -\frac{i}{2}gf_{BC}{}^A\phi^{B\ibar}\left({\Psi}^{T A\ahat}\varepsilon\overline\Upsilon{}^{\ibar}_{\ahat\bhat}\Psi^{C\bhat}+{\Psi}^{\dagger A\ahat}\varepsilon\Upsilon^{\ibar}_{\ahat\bhat}\Psi^{C\bhat*}\right)\\
&\left.-\frac{1}{4}g^2f_{BC}{}^Af_{DE}{}^A\phi^{B\ibar}\phi^{D\ibar}\phi^{C\jbar}\phi^{E\jbar}  \right).
\end{split}
\ee

\subsection{Proof of Supersymmetry}\label{SUSY}

To check that the master Lagrangian (\ref{MASTER}) is supersymmetric it suffices to show that the $\susy=1$ action (\ref{10-d action}) is invariant under the transformations (\ref{SUSYTRANS}), since the former is simply a dimensional reduction of the latter.  To prove that the action (\ref{10-d action}) is supersymmetric we will vary it explicitly, following the method found in the literature \cite{dray2000octonionic,Schray:1994fc,Baez:2009xt}. It turns out that the variation vanishes by virtue of the alternativity of the division algebras. Varying the action gives
\begin{equation}\label{10-d variation}
\begin{split}
\delta S&=\int d^{n+2}x\left(\delta A_\nu^A D_\mu F^{A\mu\nu}-\text{Re}(i\delta\Psi^{\dagger A}\sigmabar^\mu D_\mu\Psi^A+igf_{BC}{}^A\Psi^{\dagger A}\sigmabar^\mu \delta A_\mu^B\Psi^C+i\Psi^{\dagger A}\sigmabar^\mu D_\mu\delta \Psi^A)\right)\\
&=\int d^{n+2}x\left(\delta A_\nu^A D_\mu F^{A\mu\nu}-\text{Re}(igf_{BC}{}^A\Psi^{\dagger A}\sigmabar^\mu \delta A_\mu^B\Psi^C+2i\Psi^{\dagger A}\sigmabar^\mu D_\mu\delta \Psi^A)\right).
\end{split}
\end{equation}
First, we prove that the `3 $\Psi$'s' term $\text{Re}(igf_{BC}{}^A\Psi^{\dagger A}\sigmabar^\mu \delta A_\mu^B\Psi^C)$ vanishes. To see this, define the trace-reversed outer product of spinors $\Psi_1$ and $\Psi_2$ by
\be
\begin{split}
\Psi_1\cdot\Psi_2 &=i(\Psi_1\Psi_2^\dagger-\Psi_2\Psi_1^\dagger)-(\text{trace})\\
&=i(\Psi_1\Psi_2^\dagger-\Psi_2\Psi_1^\dagger)+i(\Psi_2^\dagger\Psi_1-\Psi_1^\dagger\Psi_2)\id\\
&=\text{Re}(i\Psi_1^\dagger\sigmabar_\mu\Psi_2)\sigmabar^\mu.
\end{split}
\ee
Note that the trace $-i(\Psi_2^\dagger\Psi_1-\Psi_1^\dagger\Psi_2)$ is a real number. If we then act with this on a third spinor $\Psi_3$ and add cyclic permutations we find we get zero:
\be
\begin{split}
(\Psi_1\cdot\Psi_2)\Psi_3+(\Psi_2\cdot\Psi_3)\Psi_1+(\Psi_3\cdot\Psi_1)\Psi_2&= i(\Psi_1\Psi_2^\dagger-\Psi_2\Psi_1^\dagger)\Psi_3+i(\Psi_2^\dagger\Psi_1-\Psi_1^\dagger\Psi_2)\Psi_3+\text{ cyclic permutations}\\
&=i(\Psi_1\Psi_2^\dagger-\Psi_2\Psi_1^\dagger)\Psi_3+i\Psi_3(\Psi_2^\dagger\Psi_1-\Psi_1^\dagger\Psi_2)+\text{ cyclic permutations}\\
&=i\left(+[\Psi_1,\Psi_2^\dagger,\Psi_3]+[\Psi_2,\Psi_3^\dagger,\Psi_1]+[\Psi_3,\Psi_1^\dagger,\Psi_2]\right.\\
&\left.\hspace{0.77cm}-[\Psi_3,\Psi_2^\dagger,\Psi_1]-[\Psi_1,\Psi_3^\dagger,\Psi_2]-[\Psi_2,\Psi_1^\dagger,\Psi_3]\right)\\
&=0,
\end{split}
\ee
since the associators cancel in pairs by alternativity. For example,
\be
\begin{split}
[\Psi_1,\Psi_2^\dagger,\Psi_3]-[\Psi_1,\Psi_3^\dagger,\Psi_2]&=\Psi_{1i}\Psi_{2j}^T\Psi_{3k}[e_i,e_j^*,e_k]-\Psi_{1i}\Psi_{3j}^T\Psi_{2k}[e_i,e_j^*,e_k]\\
&=-\Psi_{1i}\Psi_{2j}^T\Psi_{3k}[e_i,e_j,e_k]+\Psi_{1i}\Psi_{3j}^T\Psi_{2k}[e_i,e_j,e_k]\\
&=\Psi_{1i}\Psi_{3j}^T\Psi_{2k}\left([e_i,e_k,e_j]+[e_i,e_j,e_k]\right)\\
&=0.
\end{split}
\ee
With a little more work, using the form of the variation without $\sigma$ matrices in equation (\ref{variation no sigma}), we can show that the 3 $\Psi$'s term can be rewritten
\be
\begin{split}
gf_{BC}{}^A\text{Re}\left(i\Psi^{\dagger A}\sigmabar^\mu \delta A_\mu^B\Psi^C\right)&=gf_{BC}{}^A\text{Re}\left(i\epsilon^{\dagger}(\Psi^C\cdot\Psi^A)\Psi^B\right)\\
&=\frac{1}{3}gf_{BC}{}^A\text{Re}\left(i\epsilon^{\dagger}\left[(\Psi^C\cdot\Psi^A)\Psi^B
+(\Psi^A\cdot\Psi^B)\Psi^C+(\Psi^B\cdot\Psi^C)\Psi^A\right]\right)=0,
\end{split}
\ee
where we have used the fact that $\Psi^A\cdot\Psi^B=-\Psi^B\cdot\Psi^A$. Thus we have proven that the 3 $\Psi$'s term is zero by virtue of the alternativity of the division algebras. To prove that the remaining terms in $\delta S$ are zero we substitute in the variations of $A$ and $\Psi$ and invoke the Fierz identity
\be
\sigmabar^\mu(\sigma^{[\nu}(\sigmabar^{\rho]}~\cdot~))=\sigmabar^{[\mu}(\sigma^{\nu}(\sigmabar^{\rho]}~\cdot~))+2\eta^{\mu[\nu}(\sigmabar^{\rho]}~\cdot~).
\ee 
This gives
\begin{equation}\label{10-d variation2}
\begin{split}
\delta S&=\int d^{n+2}x\left(\text{Re}(i\Psi^{A\dagger}\sigmabar_\nu \epsilon) D_\mu F^{A\mu\nu}-\text{Re}(i\Psi^{\dagger A}\sigmabar^\mu D_\mu(\hat{F}^A\epsilon))\right)\\
&=\int d^{n+2}x\left(\text{Re}(i\Psi^{A\dagger}\sigmabar_\nu \epsilon) D_\mu F^{A\mu\nu}-\frac{1}{2}\text{Re}(i\Psi^{\dagger A}\sigmabar^\mu (\sigma^{[\nu}(\sigmabar^{\rho]}\epsilon))D_\mu F^A_{\nu\rho}-\frac{1}{2}\text{Re}(i\Psi^{\dagger A}\sigmabar^\mu (\sigma^{[\nu}(\sigmabar^{\rho]}\partial_\mu\epsilon)) F^A_{\nu\rho}\right)\\
&= \int d^{n+2}x\left(-\frac{1}{2}\text{Re}(i\Psi^{\dagger A}\sigmabar^{[\mu} (\sigma^{\nu}(\sigmabar^{\rho]}\epsilon))D_{[\mu} F^A_{\nu\rho]}-\frac{1}{2}\text{Re}(i\Psi^{\dagger A}\sigmabar^\mu (\sigma^{[\nu}(\sigmabar^{\rho]}\partial_\mu\epsilon)) F^A_{\nu\rho}\right)\\
&=0,
\end{split}
\end{equation}
since the first term vanishes by the gauge Bianchi identity and the second (which contains the supercurrent) because $\epsilon$ is (of course) constant. The term containing the supercurrent can be rewritten (by taking the dagger and repeatedly applying (\ref{RE})) as
\be
\begin{split}
-\frac{1}{2}\text{Re}(i\Psi^{\dagger A}\sigmabar^\mu (\sigma^{[\nu}(\sigmabar^{\rho]}\partial_\mu\epsilon)) F^A_{\nu\rho}&=\frac{1}{2}\text{Re}(i((\partial_\mu\epsilon^\dagger\sigmabar^{[\rho})\sigma^{\nu]})\sigmabar^\mu\Psi^A) F^A_{\nu\rho}\\
&=\frac{1}{2}\text{Re}(i\partial_\mu\epsilon^\dagger\sigmabar^{[\rho}(\sigma^{\nu]}(\sigmabar^\mu\Psi^A))) F^A_{\nu\rho},
\end{split}
\ee
from which we read off the supercurrent
\be
\mathcal{J}^\mu=\hat{\bar{F}}^A(\sigmabar^\mu\Psi^A).
\ee

\subsection{Supersymmetry Algebra and Off-Shell Formulation}

Remaining with the $\susy=1$ theories in $D=n+2$ dimensions, we can take the commutators of the supersymmetry transformations given in equation (\ref{SUSYTRANS}):
\be
\begin{split}
[\delta_1,\delta_2]A_\mu^A&=\text{Re}(i\epsilon_2^\dagger\sigmabar^\nu\epsilon_1) F^A_{\nu\mu},\\
[\delta_1,\delta_2]\Psi^A&=\text{Re}(i\epsilon_2^\dagger\sigmabar^\mu\epsilon_1) D_\mu\Psi^A+\left(\frac{i}{2}[\epsilon_1,\epsilon_2^\dagger,(\sigmabar^\mu D_\mu\Psi^A)]+\frac{1}{2}\epsilon_1\text{Im}(i\epsilon_2^\dagger(\sigmabar^\mu D_\mu\Psi^A))-(1\leftrightarrow 2)\right).
\end{split}
\ee
As usual, the commutator of two supersymmetry transformations is a gauge-covariant translation, but the algebra fails to close by terms proportional to the Dirac equation $\sigmabar^\mu D_\mu\Psi^A=0$. The difference between the number of fermionic and bosonic degrees of freedom is
\be
2n-(n+1)=n-1,
\ee
so if we are to close the algebra off-shell, the counting suggests we use an auxiliary $\text{Im}(\Al)$-valued scalar field $D^A=D^A_i e_i$. This idea was explored in \cite{Berkovits:1993hx,Evans:1994cn,Baulieu:2007ew}. We add to the action (\ref{10-d action}) the term
\be
S_{\text{aux}}=\int d^{n+2}x\left(\frac{1}{2} D^{A*}D^A\right),
\ee
and modify the supersymmetry transformations to
\be
\begin{split}\label{SUSYAUX}
\delta A_\mu^A&=\text{Re}(i\Psi^{A\dagger}\sigmabar_\mu \epsilon),\\
\delta\Psi^A&=\frac{1}{2}(\hat{F}^A\epsilon+\epsilon D^A),\\
\delta D^A&=\text{Im}((iD_\mu\Psi^{A\dagger}\sigmabar_\mu) \epsilon)
\end{split}
\ee
(note that $\text{Im}(iz)=i\text{Re}(z)$ for some division algebra element $z$). It is straightforward to check that the action $S+S_{\text{aux}}$ is invariant under these transformations. However, in the $D=10$ case, multiplying the octonionic objects $\epsilon$ and $D^A$ in the transformations actually breaks the Lorentz symmetry of the theory. This is clear if we try to Lorentz transform the $\delta\Psi^A$ transformation; we can only make sense of this \cite{Evans:1994cn} if we restrict from $\SO(1,9)$ to $\SO(1,2)\times G_2$ and allow $D^A$ to transform as the $\bold{7}$ of $G_2$. This is a result of the fact that left- and right-multiplication do not commute in $\Oct$, due to non-associativity. We conclude that in the $D=10$ octonionic theory the imaginary auxiliary field is not a scalar at all but a $G_2$ vector. By a similar argument the auxiliary field in $D=6$ remains a scalar under Lorentz transformations but transforms as a $\bold{3}$ under the R-symmetry Sp(1).

The commutators of the supersymmetry transformations (\ref{SUSYAUX}) are as follows:
\be
\begin{split}
[\delta_1,\delta_2]A_\mu^A&=\text{Re}(i\epsilon_2^\dagger\sigmabar^\nu\epsilon_1) F^A_{\nu\mu}
-\frac{1}{2}\text{Re}\left(i[\epsilon_2^\dagger,\sigmabar_\mu,\epsilon_1]D^A\right),\\
[\delta_1,\delta_2]\Psi^A&=\text{Re}(i\epsilon_2^\dagger\sigmabar^\mu\epsilon_1) D_\mu\Psi^A+\left(\frac{i}{2}[\epsilon_1,\epsilon_2^\dagger,(\sigmabar^\mu D_\mu\Psi^A)]-(1\leftrightarrow 2)\right),\\
[\delta_1,\delta_2]D^A&=\text{Re}(i\epsilon_2^\dagger\sigmabar^\mu\epsilon_1) D_\mu D^A
-i[\epsilon_2^\dagger,\sigmabar^\nu,\epsilon_1]D^\mu F^A_{\mu\nu}+\frac{i}{2}\text{Re}\left([\epsilon_2^\dagger,\sigmabar^\mu,\epsilon_1]D_\mu D^A\right).
\end{split}
\ee
The algebra is closed for the associative algebras $\R,\C,\Q,$ corresponding to $D=3,4,6,$ but fails to close by associators for $D=10$ over $\Oct$. Interestingly, in the octonionic case all of these associators vanish if we set 7 of the 16 real components of the supersymmetry parameters to zero by constraining one of the two octonionic components of $\epsilon$ to be real:
\be\label{EPSREAL}
\epsilon=\begin{pmatrix} \epsilon_1 \\ \epsilon_2 \end{pmatrix} \rightarrow \begin{pmatrix} \text{Re}(\epsilon_1) \\ \epsilon_2 \end{pmatrix}
\ee
(note that in this equation the subscripts 1 and 2 refer to the components of a single supersymmetry parameter $\epsilon$, while in the algebra equations above the same subscripts label two different supersymmetry parameters). This is in agreement with \cite{Berkovits:1993hx,Evans:1994cn}, where an imaginary octonionic auxiliary field was used to close the algebra for 9 out of 16 supersymmetries.

Since the master Lagrangian comes from dimensional reduction of the $\susy=1$ theories, it is clear that the appropriate auxiliary field for a theory with extended supersymmetry should be valued in $\text{Im}(\Al_{n\susy})$; otherwise, the form of the new terms in the supersymmetry transformations (\ref{SUSYAUX}) is unchanged. Interestingly, the transformations in the $D=3$ octonionic case ($n=1$, $\susy=8$) are Lorentz-covariant with symmetry $\SO(1,2)\times G_2$. However, this must be broken to $\SO(1,1)\times G_2$ to close the algebra \cite{Evans:1994cn}, since one must impose the constraint of equation (\ref{EPSREAL}).

\section{Summary and Further Work}\label{CONC}

We have seen that it is possible to write any super Yang-Mills theory in
$D=n+2$ using a pair of division algebras: $\Al_{n\susy}$ and $\Al_n$. We
also established the role of triality algebras in these theories; the total
symmetry $\widetilde{\mathfrak{tri}}(\Al_{n\susy},\Al_n)$ of the on-shell
degrees of freedom is the subgroup of $\mathfrak{tri}(\Al_{n\susy})$ for
which one element of the triality triple preserves a subalgebra $\Al_n\subseteq\Al_{n\susy}$.
 Finally, we used imaginary $\Al_{n\susy}$-valued auxiliary fields to close
the non-maximal supersymmetry algebra off-shell. The failure to close for
maximally supersymmetric theories is attributed directly to the non-associativity
of the octonions.

A previous paper \cite{Borsten:2013bp} tensored the multiplets of the $D=3$
Yang-Mills theories over $\R,\C,\Q,\Oct$ to obtain supergravity theories
whose U-dualities fill out the magic square. The natural next step is to
generalise this to $D=3,4,6,10$ and obtain a `magic pyramid' with the $D=3$
magic square at the base and Type II supergravity at the summit.

On the basis of these results we speculate that the part played by octonions in string and M-theory may be more prominent than previously thought.

\appendix

\section{SO(8) Triality}\label{APPA}

To study triality it is useful to talk in terms of derivations. A derivation of an algebra $\Al$ is an $\R$-linear map $d:\Al\rightarrow\Al$ such that
\be\label{DERIVATIONS}
d(xy)=d(x)y+xd(y),~~~~x,y\in\Al.
\ee
It can be shown \cite{Schafer:1966} that when $\Al$ is a division algebra every derivation of an element $z\in\Al$ is of the form
\be
\hat{d}_{x,y}(z)\equiv\frac{1}{2}[x,y,z]-\frac{1}{6}\left[[x,y],z\right],
\ee
for some $x,y\in\Al$. We define the derivations operators
\be
\hat{G}_{ij}\equiv \frac{1}{2}[e_i,e_j,~\cdot~]-\frac{1}{6}\left[[e_i,e_j],~\cdot~\right].
\ee
When $\Al=\Oct$, the derivations $\hat{G}_{ij}$ are in fact the generators of $G_2$, the automorphisms of the octonions. That is, if we consider exponentiating the derivations to obtain $G_2$ group elements $g$, then the finite version of equation (\ref{DERIVATIONS}) is
\be
g(xy)=g(x)g(y),
\ee
so $G_2$ transformations preserve the octonionic multiplication rule. It is useful to note the effect of acting with $\hat{G}_{ij}$ on an octonion $z$ with components $z_a$:
\be
\hat{G}_{ij}(z)=P^{\bold{14}}_{ijkl}J_{[kl]mn}z_ne_m,
\ee
where $J_{[ij]}$ are the usual generators of SO(7) and $P^{\bold{14}}$ is a projection operator given by
\be
P^{\bold{14}}_{ijkl}=\frac{2}{3}\left(\delta_{i[k}\delta_{l]j}-\frac{1}{4}Q_{ijkl}\right),~~~~~~P^{\bold{14}}_{ijkl}P^{\bold{14}}_{klmn}=P^{\bold{14}}_{ijmn}.
\ee
The $G_2$ transformation thus only affects the imaginary components of an octonion, and acts as the subgroup of SO(7) projected onto by $P^{\bold{14}}$, a rank 14 `matrix' on the 21-dimensional vector space of antisymmetric $7\times7$ matrices (so $G_2$ is 14-dimensional).

Equations (\ref{tri1}), (\ref{tri2}) and (\ref{tri3}) are the essential ingredient of SO(8) triality. Let us focus on the first of these three definitions, $a=\chi\psi^*$, and consider acting with an infinitesimal SO(8) transformation. With a little algebra we can show that the spinor and vector transformations may be written
\be
\begin{split}
\delta\psi &= \frac{1}{4}\theta^{ab}e_a^*(e_b\psi)=\hat{d}(\psi)+\psi\alpha-\beta\psi-\psi\beta,\\
\delta\chi &= \frac{1}{4}\theta^{ab}e_a(e_b^*\chi)=\hat{d}(\chi)+\alpha\chi+\chi\alpha-\chi\beta,\\
\delta a&=\frac{1}{4}\theta^{ab}\Big(e_a(e_b^*a)-a(e_a^*e_b)\Big)=\hat{d}(a)+\alpha a + a\beta,
\end{split}
\ee
where
\be
\hat{d}= \frac{1}{2}\theta^{ij}\hat{G}_{ij},~~~~~ \alpha=\left(-\frac{1}{2}\theta^{0i}-\frac{1}{12}C_{ijk}\theta^{jk}\right)e_i, ~~~~~ \beta = \left(-\frac{1}{2}\theta^{0i}+\frac{1}{12}C_{ijk}\theta^{jk}\right)e_i.
\ee
This corresponds to the decomposition
\be
\SO(8)\supset G_2; \quad
\bold{28}\rightarrow \bold{14}+\bold{7}+\bold{7},
\ee
with the $\bold{14}$ corresponding to derivations and the two $\bold{7}$s corresponding to left- and right-multiplication by imaginary octonions. We can see what this means for the vector in terms of the usual SO(8) generators by means of more projection operators:
\be
\begin{split}
\alpha a &=\frac{1}{2}\theta^{ab}P^{\bold{7_L}}_{abcd}J_{[cd]ef}a_fe_e =\theta^{ab}P^{\bold{7_L}}_{abcd}e_ca_d ,\\
a\beta  &=\frac{1}{2}\theta^{ab}P^{\bold{7_R}}_{abcd}J_{[cd]ef}a_fe_e =\theta^{ab}P^{\bold{7_R}}_{abcd}e_ca_d ,
\end{split}
\ee
where we define
\be
\begin{split}
P^{\bold{7_L}}_{abcd}&=\frac{1}{6}\left(\delta_{a[c}\delta_{d]b}+\frac{1}{2}Q_{abcd}+4\delta_{0[a}\delta_{b][d}\delta_{c]0}\right) +\frac{1}{2}\delta_{0[a}C_{b]cd}+\frac{1}{6}\delta_{0[c}C_{d]ab},\\
P^{\bold{7_R}}_{abcd}&=\frac{1}{6}\left(\delta_{a[c}\delta_{d]b}+\frac{1}{2}Q_{abcd}+4\delta_{0[a}\delta_{b][d}\delta_{c]0}\right) -\frac{1}{2}\delta_{0[a}C_{b]cd}-\frac{1}{6}\delta_{0[c}C_{d]ab},\\
P^{\bold{14}}_{abcd}&=\frac{2}{3}\left(\delta_{a[c}\delta_{d]b}-\frac{1}{4}Q_{abcd}-2\delta_{0[a}\delta_{b][d}\delta_{c]0}\right),
\end{split}
\ee
and they satisfy
\be
\begin{split}
P^{\bold{X}}_{abef}P^{\bold{Y}}_{efcd}=P^{\bold{X}}_{abcd} \delta_{\bold{XY}}, ~~~~~ \bold{X},\bold{Y}=\bold{14},\bold{7_L},\bold{7_R}, \\
P^{\bold{14}}_{abcd}+P^{\bold{7_L}}_{abcd}+P^{\bold{7_R}}_{abcd}=\delta_{c[a}\delta_{b]d}.~~~~~~
\end{split}
\ee

Now, temporarily setting the $\alpha$ and $\beta$ parts to zero and transforming $a=\chi\psi^*$ we find that
\be
\delta a = \delta\chi\psi^*+\chi\delta\psi^*=\hat{d}(\chi) \psi^*+\chi \hat{d}(\psi^*)=\hat{d}(\chi\psi^*)=\hat{d}(a),
\ee
exactly as required (here we used the fact that $\hat{d}(\psi)^*=\hat{d}(\psi^*)$). Similar results hold when we transform using only the $\alpha$ or $\beta$ parts. The conclusion is that the tranformation rules of the vector, spinor and conjugate spinor are exactly such that the three definitions (\ref{tri1}), (\ref{tri2}) and (\ref{tri3}) make sense. We summarise the transformation rules in Table \ref{tab:triality}, where $L_\alpha$ ($R_\alpha$) represents left-(right-)multiplication by $\alpha$.

\begin{table}
\begin{tabular}{ cc | c  c cc c c cc ccccccccccc }
\hline  
\hline                     
 &&&& $\bold{14}$ &&&&& $\bold{7_L}$  &&&&& $\bold{7_R}$  \\
\hline
     $\bold{8}_v$ &&&& $d$ &&&&& $L_\alpha$   &&&&& $R_\beta$    \\
  $\bold{8}_s$ &&&& $d$ &&&&& $R_\alpha$  &&&&& $-L_\beta-R_\beta$   \\
  $\bold{8}_c$ &&&& $d$ &&&&& $L_\alpha+R_\alpha$   &&&&&$-R_\beta$  \\
\hline  
\hline
\end{tabular}
\caption{\footnotesize{A summary of SO(8) transformations for the vector, spinor and conjugate spinor. We decompose the adjoint of SO(8) into $G_2$ irreps, and their action on the $\bold{8}_v,\bold{8}_s,\bold{8}_c$ is given by the entries in the table. \label{tab:triality}}}
\end{table}

There is a discrete triality symmetry (isomorphic to the group $S_3$ of permutations on three objects) given by permutation of the three representations. These outer automorphisms are reflected in the manifest symmetry of the SO(8) Dynkin diagram. The actual permutations must be taken carefully as even permutations of $\{a,\psi,\chi^*\}$ or odd permutations of $\{a^*,\psi^*,\chi\}$. For example, equations (\ref{tri2}) and (\ref{tri3}) may be obtained from equation (\ref{tri1}) by cyclically permuting $\{a,\psi,\chi^*\}$. In other words we can think of the elements of the group of outer automorphisms of SO(8) (or, more correctly, Spin(8)) as being the following permutations mixed with octonionic conjugations:
\be
\begin{array}{llllll}
\begin{pmatrix} a & \psi & \chi^* \\ a & \psi & \chi^* \end{pmatrix}, &
\begin{pmatrix} a & \psi & \chi^* \\ \psi & \chi^* & a  \end{pmatrix},&
\begin{pmatrix} a & \psi & \chi^* \\ \chi^*& a & \psi   \end{pmatrix}, \\[10pt]
\begin{pmatrix} a & \psi & \chi^* \\ \chi & \psi^* & a^* \end{pmatrix},&
\begin{pmatrix} a & \psi & \chi^* \\  \psi^* & a^*& \chi \end{pmatrix}, &
\begin{pmatrix} a & \psi & \chi^* \\   a^*& \chi &\psi^* \end{pmatrix}.
\end{array}
\ee
It is clear that these operations leave equations (\ref{tri1}), (\ref{tri2}) and (\ref{tri3}) invariant, as well as Table \ref{tab:triality2}.

\begin{figure}[h!]
  \centering
    \includegraphics[width=0.15\textwidth]{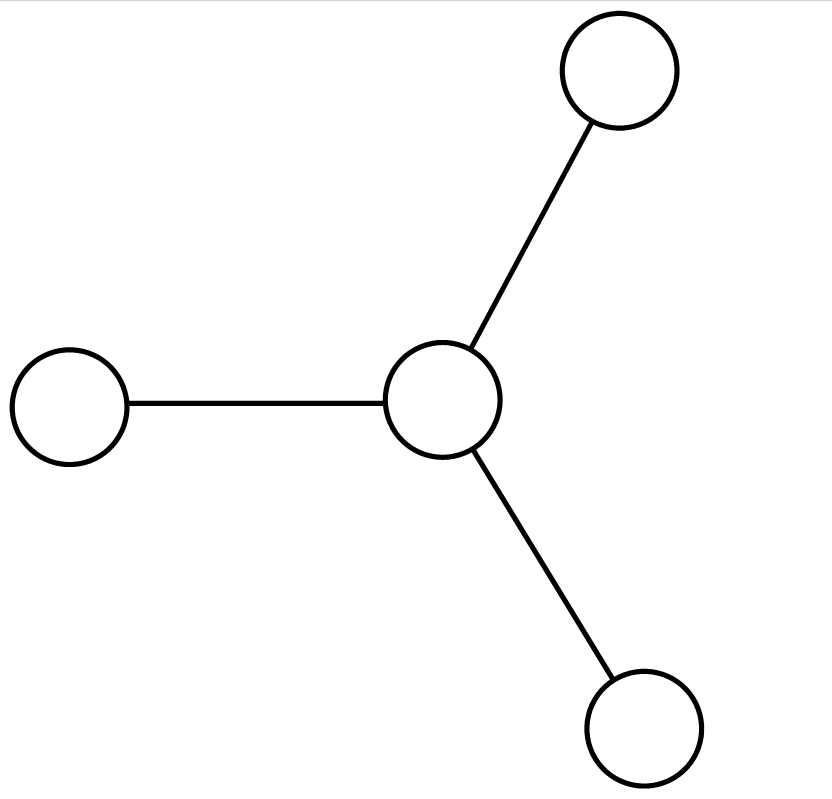}
  \caption{\footnotesize{The Dynkin diagram of SO(8)}}
\end{figure}
\begin{table}[t!]
\begin{tabular}{ cc | c  c cc c c cc cccccccccc }
\hline            
\hline
 &&&& $\bold{14}$ &&&&& $\bold{7_L}$  &&&&& $\bold{7_R}$ &&&&& $-\bold{7_L}-\bold{7_R}$ \\
\hline
     $\bold{8}_v$ &&&& $d$ &&&&& $L_\alpha$   &&&&& $R_\beta$   &&&&&  $-L_\gamma-R_\gamma$ \\
  $\bold{8}_s$ &&&& $d$ &&&&& $R_\alpha$  &&&&& $-L_\beta-R_\beta$ &&&&& $L_\gamma$    \\
  $\bold{8}_c^*$ &&&& $d$ &&&&& $-L_\alpha-R_\alpha$   &&&&&$L_\beta$  &&&&& $R_\gamma$  \\
\hline
\hline
\end{tabular}
\caption{\footnotesize{SO(8) transformations on $\{a,\psi,\chi^*\}$. A third column of transformations, obtained as a linear combination of the first two (with $\gamma$ some imaginary octonion), has been added in order to emphasise the cyclic symmetry between the three reps. We can also make odd permutations if we simultaneously take the octonionic conjugate of every entry in the table. \label{tab:triality2}}}
\end{table}
We can now rewrite the table of SO(8) transformations with the $\bold{8}_c$ octonionic-conjugated (denoted $\bold{8}_c^*$) and we see that the cyclic symmetry between $a$, $\psi$ and $\chi^*$ becomes manifest - see Table \ref{tab:triality2}. To make an odd permutation we must take the octonionic conjugate of every entry in the table (where $L_\alpha^*=-R_\alpha$).

%\bibliographystyle{utphys}
%\bibliography{Ref_Library}

\providecommand{\href}[2]{#2}\begingroup\raggedright\endgroup

\end{document}